\documentclass[camera]{jpaper}
\usepackage[nocompress]{cite}

\usepackage{soul}

\usepackage{amsmath}
\usepackage{graphicx}
\usepackage[linesnumbered,ruled]{algorithm2e}

\usepackage{algpseudocode}
\usepackage{titlesec}
\algrenewcommand\algorithmiccomment[2][\normalsize]{{#1\hfill\(\triangleright\) #2}}
\usepackage{geometry} \titlespacing*{\section}{0pt}{3pt}{-1pt}
\titlespacing*{\subsection}{0pt}{3pt}{1pt}
\titlespacing*{\subsubsection}{0pt}{1pt}{0pt}

\usepackage{array}
\makeatletter
\let\MYcaption\@makecaption
\makeatother

\usepackage[font=footnotesize]{subcaption}

\makeatletter
\let\@makecaption\MYcaption
\makeatother

\usepackage{fixltx2e}

\usepackage{dblfloatfix}
\usepackage[nolessnomore, italic]{mathastext}
\usepackage[T1]{fontenc}
\usepackage[usenames,dvipsnames,svgnames,table]{xcolor}
\usepackage{multirow}
\usepackage{hhline}
\usepackage[normalem]{ulem}
\usepackage{setspace}
\usepackage{indentfirst}
\usepackage{footmisc}

\usepackage{pifont}
\newcommand{\cmark}{\ding{51}}%
\newcommand{\xmark}{\ding{55}}%

\newcommand{\squeezeme}{ \setlength{\itemsep}{0pt}
     \setlength{\parsep}{3pt}
     \setlength{\topsep}{3pt}
     \setlength{\partopsep}{0pt}
     \setlength{\leftmargin}{1.5em}
     \setlength{\labelwidth}{1em}
     \setlength{\labelsep}{0.5em} }

\newcommand{\incircle}[1]{\raisebox{0.5pt}{\protect \textcircled{\raisebox{-0.8pt}{#1}}}}

\newif\ifcameraready
\camerareadytrue

\ifcameraready

\newcommand{\affilCMU}[0]{\textsuperscript{$\ddagger$}}
\else

\newcommand{\affilCMU}[0]{\textsuperscript{$\dagger$}}
\fi
\newcommand{\affilETH}[0]{\textsuperscript{\S}}

\definecolor{amber}{rgb}{1.0, 0.49, 0.0}
\definecolor{darkgreen}{rgb}{0.0, 0.2, 0.13}
\definecolor{darkbyzantium}{rgb}{0.36, 0.22, 0.33}
\definecolor{darkseagreen}{rgb}{0.56, 0.74, 0.56}
\definecolor{darkspringgreen}{rgb}{0.09, 0.45, 0.27}
\definecolor{dollarbill}{rgb}{0.52, 0.73, 0.4}

\newcommand{\mechanism}[0]{D-RaNGe}
\newcommand{\jk}[1]{{\color{black}#1}}
\newcommand{\jktwo}[1]{{\color{black}#1}}
\newcommand{\jkthree}[1]{{\color{black}#1}}
\newcommand{\jkfour}[1]{{\color{black}#1}}
\newcommand{\jkfive}[1]{{\color{black}#1}}

\newcommand{\hasan}[1]{{\color{black}#1}}
\newcommand{\hh}[1]{{\color{black}#1}}
\newcommand{\hhtwo}[1]{{\color{black}#1}}
\newcommand{\hhthree}[1]{{\color{black}#1}}
\newcommand{\hhf}[1]{{\color{black}#1}}
\newcommand{\changes}[1]{{\color{black}#1}}
\newcommand{\delete}[1]{{\color{red}\sout{}}}

\newcommand{\mpx}[1]{{\color{black}#1}}
\newcommand{\mpy}[1]{{\color{black}#1}}
\newcommand{\mpo}[1]{{\color{black}#1}}
\newcommand{\mpt}[1]{{\color{black}#1}}
\newcommand{\mpf}[1]{{\color{black}#1}}

\begin{document}
\bstctlcite{IEEEexample:BSTcontrol} 


\title{\mechanism: Using Commodity DRAM Devices \\ to Generate True Random Numbers \\ with Low Latency and High Throughput}


%


\author{
{Jeremie S. Kim\affilCMU\affilETH}\qquad%
{Minesh Patel\affilETH}\qquad%
{Hasan Hassan\affilETH}\qquad%
{Lois Orosa\affilETH}\qquad%
{Onur Mutlu\affilETH\affilCMU}\qquad\\%
{\affilCMU Carnegie Mellon University \qquad \affilETH ETH Z{\"u}rich}%
\vspace{-5pt}%
}


%


\maketitle
\thispagestyle{plain} 
\pagestyle{plain}

\setstretch{0.82}
\renewcommand{\footnotelayout}{\setstretch{0.9}}







%


\begin{abstract}

We propose a new DRAM-based true random number generator (TRNG) that
leverages DRAM cells as an entropy source. \hhtwo{The key idea is to}
intentionally \hhtwo{violate} the \hhtwo{DRAM} access timing parameters and use
the resulting errors as the source of randomness. Our technique specifically
decreases the DRAM row activation latency (timing parameter $t_{RCD}$) below
manufacturer-recommended specifications, to induce read errors, or activation
failures, that exhibit true random behavior. We then aggregate the resulting
data from multiple cells to obtain a TRNG capable of providing a
\jkfour{high throughput} of random numbers at low latency.

To demonstrate that our TRNG design is viable using commodity DRAM chips, we
rigorously characterize the behavior of activation failures in 282
state-of-the-art LPDDR4 \mpt{devices from three major DRAM manufacturers. We
verify our observations using four additional DDR3 DRAM devices from the same
manufacturers}. Our results show that many cells in each device produce random data
that remains robust over both time and temperature variation. We use our
observations to develop \mechanism, a methodology for extracting true random
numbers from commodity DRAM devices with high throughput and low latency by
deliberately violating \hhtwo{the} read access timing parameters. We evaluate the quality
of our TRNG using the commonly-used NIST statistical test suite for randomness
and find that \mechanism: 1) successfully passes each test, and 2) generates
true random numbers with over two orders of magnitude higher throughput
than the previous highest-throughput DRAM-based TRNG. 

\end{abstract}

\section{Introduction} 

Random number generators (RNGs) are critical components in many different
applications, including cryptography, scientific simulation, industrial
testing, and recreational entertainment~\mpt{\cite{bagini1999design,
rock2005pseudorandom, ma2016quantum, stipvcevic2014true,
barangi2016straintronics, tao2017tvl, gutterman2006analysis, von2007dual,
kim2017nano, drutarovsky2007robust, kwok2006fpga, cherkaoui2013very,
zhang2017high, quintessence2015white}}. These applications require a mechanism
capable of rapidly generating random numbers across \mpt{a wide variety of
operating conditions (e.g., temperature/voltage fluctuations, manufacturing
variations, malicious external attacks)~\cite{yang2016all}}. In particular, for
modern cryptographic applications, a random (i.e., completely
unpredictable) number generator is critical to prevent information leakage to
a potential adversary~\cite{kocc2009cryptographic, gutterman2006analysis,
von2007dual, kim2017nano, drutarovsky2007robust, kwok2006fpga,
cherkaoui2013very, zhang2017high, quintessence2015white}.

Random number generators can be broadly classified into two
categories~\cite{tilborgencyclopedia, chevalier1974random, knuth1998art,
tsoi2003compact}: 1) \emph{pseudo-random number generators
(PRNGs)}~\cite{matsumoto1998mersenne, blum1986simple, mascagni2000algorithm,
steele2014fast, marsaglia2003xorshift}, which deterministically generate
numbers starting from a \emph{seed value} with the goal of approximating a true
random sequence, and 2) \emph{true random number generators
(TRNGs)}~\cite{pyo2009dram, keller2014dynamic, sutar2018d, hashemian2015robust,
tehranipoor2016robust, wang2012flash, ray2018true, holcomb2007initial,
holcomb2009power, van2012efficient, chan2011true, tzeng2008parallel,
teh2015gpus, majzoobi2011fpga, wieczorek2014fpga, chu1999design,
amaki2015oscillator, mathew20122, brederlow2006low, tokunaga2008true,
bucci2003high, bhargava2015robust, kinniment2002design, holleman20083,
gutterman2006analysis, dorrendorf2007cryptanalysis, lacharme2012linux,
pareschi2006fast, yang2016all}, which generate random numbers based on
sampling non-deterministic random variables inherent in various physical
phenomena (e.g., electrical noise, atmospheric noise, clock jitter, Brownian
motion). 

PRNGs are popular due to their flexibility, low cost, and fast pseudo-random
number generation time~\cite{chan2011true}, but their output is \emph{fully
determined} by the starting seed value. This means that the output of a PRNG
may be predictable given complete information about its operation.
Therefore, a PRNG falls short for applications that require high-entropy
values~\cite{corrigan2013ensuring, von2007dual, cherkaoui2013very}.  In
contrast, because a TRNG mechanism relies on sampling entropy inherent in
\emph{non-deterministic} physical phenomena, the output of a TRNG is
\emph{fully unpredictable} even when complete information about the underlying
mechanism is available~\cite{kocc2009cryptographic}.




Based on analysis done by prior work on TRNG
design~\cite{kocc2009cryptographic, jun1999intel, schindler2002evaluation},
we argue that an \emph{effective} TRNG must: \hhthree{1) produce truly
random (i.e., completely unpredictable) numbers, 2) provide a high
throughput of random numbers at low latency, and 3) be practically
implementable at low cost.} Many prior works study different methods of
generating true random numbers that can be implemented using CMOS
devices~\cite{pyo2009dram, keller2014dynamic, sutar2018d,
hashemian2015robust, tehranipoor2016robust, wang2012flash, ray2018true,
holcomb2007initial, holcomb2009power, van2012efficient, chan2011true,
tzeng2008parallel, teh2015gpus, majzoobi2011fpga, wieczorek2014fpga,
chu1999design, amaki2015oscillator, mathew20122, brederlow2006low,
tokunaga2008true, bucci2003high, bhargava2015robust, kinniment2002design,
holleman20083, gutterman2006analysis, dorrendorf2007cryptanalysis,
lacharme2012linux, pareschi2006fast, yang2016all}. We provide a thorough
discussion of these past works in Section~\ref{related}.
\mpt{Unfortunately, most of these proposals fail to satisfy all of the
properties of an effective TRNG because they} either require specialized
hardware to implement (e.g., free-running
oscillators~\cite{amaki2015oscillator, yang2016all}, metastable
circuitry~\cite{mathew20122, brederlow2006low, tokunaga2008true,
bhargava2015robust}) or are unable to sustain continuous high-throughput
operation on the order of Mb/s (e.g., memory startup
values~\cite{tehranipoor2016robust, holcomb2007initial, holcomb2009power,
van2012efficient, eckert2017drng}, memory data retention
failures~\cite{keller2014dynamic, sutar2018d}). These limitations preclude the
widespread adoption of such TRNGs, thereby limiting the overall impact of these
proposals.

Commodity DRAM chips offer a promising substrate to overcome these
limitations due to three major reasons. First, DRAM
operation is highly sensitive to changes in access timing, which means
that we can easily induce failures by manipulating manufacturer-recommended
DRAM access timing parameters. \jkfour{These failures have been shown to exhibit
non-determinism~\cite{yaney1987meta, nair2016xed, khan2014efficacy,
chang2016understanding, qureshi2015avatar, patel2017reaper, kim2018dram,
lee2015adaptive, lee-sigmetrics2017, kim2018solar} and therefore \jkfive{they}
may be exploitable for true random number generation.} Second, commodity DRAM
devices already provide an interface capable of transferring data continuously
with high throughput in order to support a high-performance TRNG. Third, DRAM
devices are already prevalently in use throughout modern computing systems,
ranging from simple microcontrollers to sophisticated supercomputers. 

\textbf{Our goal} in this paper is to design a TRNG
that:
\begin{enumerate}
\squeezeme
\item is implementable on commodity DRAM devices today
\item is fully non-deterministic (i.e., it is impossible to predict the next output even with complete information about the underlying mechanism)
\item provides continuous (i.e., constant rate), high-throughput random values \hhtwo{at low latency}
\item provides random values while minimally affecting concurrently-running applications
\end{enumerate}
Meeting these four goals would enable a TRNG design that is suitable
for applications requiring high-throughput true random number generation in
commodity devices today.

Prior approaches to DRAM-based TRNG design successfully use DRAM data retention
failures~\cite{keller2014dynamic, sutar2018d, hashemian2015robust}, DRAM
startup values~\cite{tehranipoor2016robust, eckert2017drng}, and non-determinism in DRAM
command scheduling~\cite{pyo2009dram} to generate true random numbers.
Unfortunately, these approaches do not fully satisfy our four goals because
they either \hhtwo{do not exploit a fundamentally non-deterministic entropy source
(e.g., DRAM command scheduling~\cite{pyo2009dram}) or are too slow for
continuous high-throughput operation (e.g., DRAM data retention
failures~\cite{keller2014dynamic, sutar2018d, hashemian2015robust}, DRAM
startup values~\cite{tehranipoor2016robust, eckert2017drng})}.
Section~\ref{comparison} provides a detailed comparative analysis of these
prior works.

In this paper, we propose a new way to leverage DRAM cells as an entropy source
for true random number generation by intentionally violating the access
timing parameters and using the resulting errors as the source of randomness.
Our technique specifically extracts randomness from \emph{activation
failures}, i.e., DRAM errors caused by intentionally decreasing the row
activation latency (timing parameter $t_{RCD}$) below manufacturer-recommended
specifications.  Our proposal is based on \textbf{two key observations}:
\begin{enumerate}
\squeezeme

\item \hh{Reading} certain DRAM cells with a reduced activation latency returns
\emph{true random} values. 

\item An activation failure can be induced very quickly (i.e., \emph{even
faster} than a normal DRAM row activation).

\end{enumerate}

Based on these key observations, we propose \mechanism, a new methodology for
extracting true random numbers from commodity DRAM devices with high
throughput. \mpt{\mechanism~consists of two steps: 1) identifying specific
DRAM cells that are vulnerable to activation failures using a
\emph{low-latency} profiling step and 2) generating a continuous stream (i.e.,
constant rate) of random numbers by repeatedly inducing activation failures in
the previously-identified vulnerable cells.} \mechanism~runs entirely in
software and is capable of immediately running on any commodity system that
provides the ability to manipulate DRAM timing parameters within the memory
controller~\cite{bkdg_amd2013, opteron_amd}. For most other devices, a simple
software API must be exposed without any hardware changes to the commodity DRAM
device (e.g., similarly to SoftMC~\cite{hassan2017softmc,
softmc-safarigithub}), which makes \mechanism\ suitable for implementation on
most existing systems today. 

In order to demonstrate \mechanism's effectiveness, we perform a rigorous
experimental characterization of activation failures using 282 state-of-the-art
LPDDR4~\cite{2014lpddr4} DRAM devices from three major DRAM manufacturers. We
also verify our observations using four additional DDR3~\cite{jedec2012} DRAM
devices from a single manufacturer. Using the standard NIST statistical test
suite for randomness~\cite{rukhin2001statistical}, we show that \mechanism~is
able to maintain high-quality true random number generation both over 15 days
of testing and across the entire reliable testing temperature range of our
infrastructure (55$^{\circ}$C-70$^{\circ}$C). Our results show that
\mechanism's maximum (average) throughput is \changes{$717.4 Mb/s$ ($435.7
Mb/s$) using four \hhtwo{LPDDR4} DRAM channels, which is over two orders of
magnitude} higher than that of the best prior DRAM-based TRNG.


We make the following \textbf{key contributions}:

\begin{enumerate}
\squeezeme
\item We introduce \mechanism, a new methodology for extracting true random
numbers from a commodity DRAM device at \hhtwo{high} throughput \hhtwo{and low
latency}. The key idea of \mechanism~is to use DRAM cells as entropy sources to
generate true random numbers by accessing them with a latency
that is \hhtwo{lower than}
manufacturer-recommended specifications.

\item Using experimental data from 282 state-of-the-art LPDDR4 DRAM devices
from three major DRAM manufacturers, we present a rigorous characterization of 
\hhtwo{randomness in errors induced by accessing DRAM with low latency}.
Our analysis demonstrates that \mechanism~is able
to maintain high-quality random number generation both over 15 days of testing
and across the entire reliable testing temperature range of our
infrastructure (55$^{\circ}$C-70$^{\circ}$C). We verify our observations from
this study with prior works' observations on DDR3 DRAM
devices~\cite{chang2016understanding, lee-sigmetrics2017, lee2015adaptive,
kim2018solar}. Furthermore, we experimentally demonstrate on four DDR3 DRAM
devices, from a single manufacturer, that \mechanism~is suitable for implementation
in a \hhtwo{wide} range of commodity DRAM devices.

\item We evaluate the quality of \mechanism's output bitstream using the
\mpo{standard} NIST statistical test suite for
randomness~\cite{rukhin2001statistical} and find that it successfully passes
every test.  We also compare \mechanism's performance to four previously
proposed DRAM-based TRNG designs (Section~\ref{comparison}) and show that
\mechanism\ outperforms the best prior DRAM-based TRNG design by over two orders
of magnitude in terms of maximum and average throughput.


\end{enumerate} 
 
\section{Background} 
\label{sec:background}

\hh{We} provide the necessary background on DRAM and true
random number generation that is required to understand our idea of true
random number generation \hasan{using the inherent properties of DRAM}.

\subsection{Dynamic Random Access Memory (DRAM)}
\label{subsec:dram}

We briefly \hh{describe} DRAM organization \hh{and basics}. We refer the
reader to past works~\cite{lee2015adaptive, khan2016parbor,
seshadri2016simple, hassan2016chargecache, hassan2017softmc, zhang2014half,
lee2013tiered, kim2012case, seshadri2013rowclone, chang2016understanding,
lee2016reducing, chang2017thesis, chang2016low, lee-sigmetrics2017,
seshadri2017ambit, lee2015decoupled, kim2016ramulator, kim2014flipping,
bhati2015flexible, chang2017understanding, liu2013experimental,
khan2014efficacy, qureshi2015avatar, zhang2014half, khan2017detecting,
patel2017reaper, mukundan2013understanding, chang2014improving, raidr,
kim2018dram, kim2018solar, vampire2018ghose} for more detail.

\subsubsection{DRAM System Organization}
\label{subsubsec:dram_org}

\begin{figure}[h] \centering 
    \includegraphics[width=0.85\linewidth]{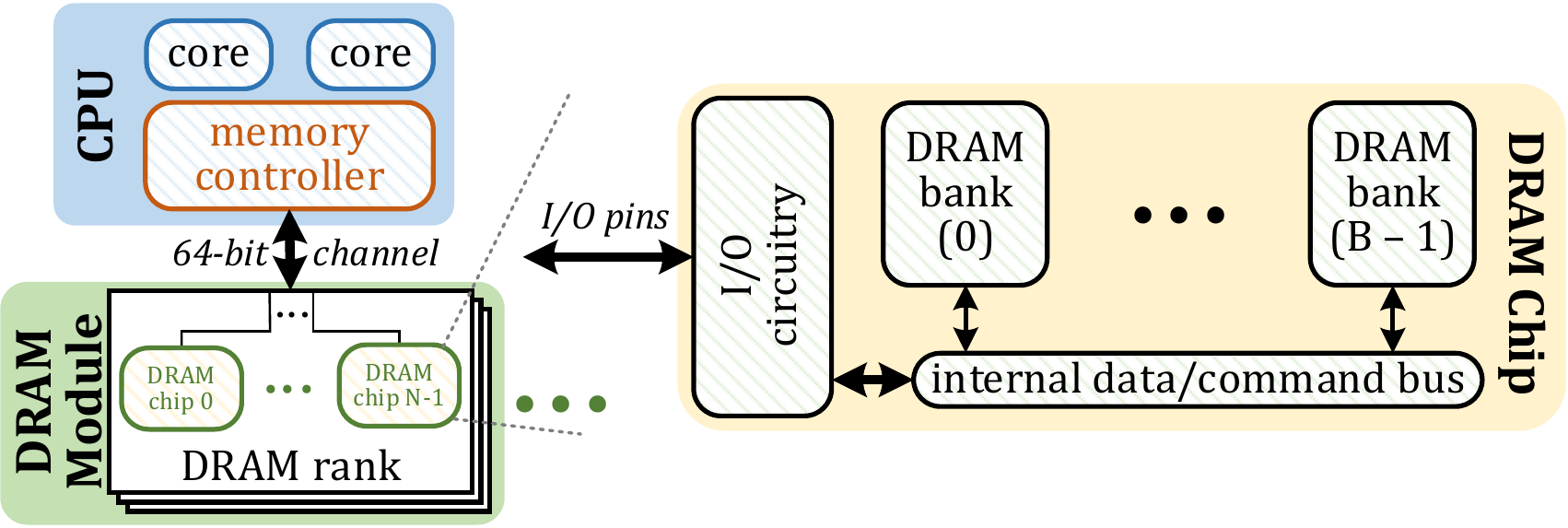} 
    \caption{A typical DRAM-based system~\cite{kim2018solar}.}
    \label{fig:dram_org}
\end{figure}

~In a typical system configuration, a CPU chip includes a set of memory
controllers, where each memory controller interfaces \hhtwo{with} a DRAM channel to
perform read and write operations. As we show in Figure~\ref{fig:dram_org}
\hh{(left)}, a DRAM channel has its own I/O bus and operates independently
of other channels in the system. \hh{To achieve high memory capacity, a
channel can host multiple DRAM modules by sharing the I/O bus between the
modules. A DRAM module implements a single or multiple DRAM ranks.}
Command and data transfers are serialized between ranks in the same channel due
to the shared I/O bus. A DRAM rank consists of multiple DRAM chips that operate
in lock-step, i.e., all chips simultaneously perform the same operation\hhtwo{,
but they do so on different bits}. The number of DRAM chips per rank depends on
the data bus width of the DRAM chips and the channel width. For example, a
typical system has a 64-bit wide DRAM channel. Thus, four 16-bit or eight 8-bit
DRAM chips are needed to build a DRAM rank.

\subsubsection{DRAM Chip Organization}
\label{subsubsec:dram_bank}

~At a high-level, a DRAM chip consists of \hh{billions} of DRAM cells that are
hierarchically organized to maximize storage density and performance. We
describe each level of the hierarchy of a modern DRAM chip.


A modern DRAM chip is composed of multiple DRAM banks (shown in
Figure~\ref{fig:dram_org}\hh{, right}). The chip communicates with the memory controller
through the \emph{I/O circuitry}. The I/O circuitry is connected to the
\emph{internal command and data bus} that is shared among all banks in the
chip.

\begin{figure}[h]
    \centering
    \begin{subfigure}[b]{.600\linewidth}
        \centering
        \includegraphics[width=\linewidth]{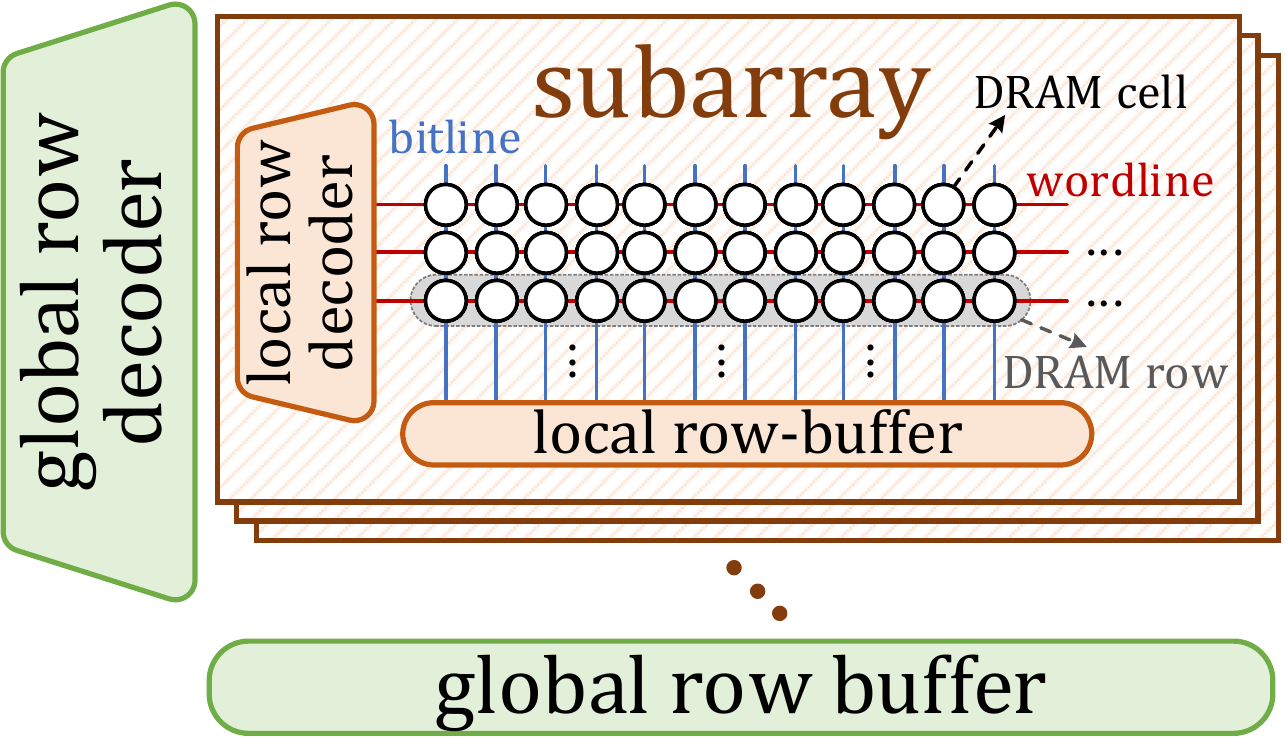}
        \caption{DRAM bank.}
        \label{subfig:dram_bank}
    \end{subfigure}
    \quad
    \begin{subfigure}[b]{.345\linewidth}
        \centering
        \includegraphics[width=\linewidth]{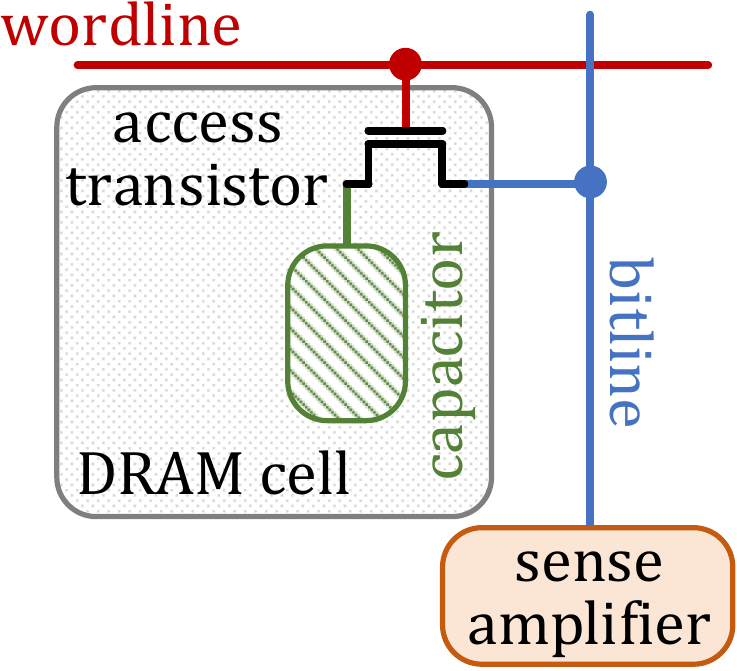}
        \caption{DRAM cell.}
        \label{subfig:dram_cell}
    \end{subfigure}

        \caption{DRAM bank and cell architecture~\cite{kim2018solar}.}
        \label{figure:dram_bank_and_cell}
\end{figure}

Figure~\ref{subfig:dram_bank} illustrates the \hasan{organization} of a DRAM bank. In a
bank, the \emph{global row decoder} partially decodes the address of the
accessed \emph{DRAM row} to select the corresponding \emph{DRAM subarray}. A
DRAM subarray is a 2D array of DRAM cells, where cells are horizontally
organized into multiple DRAM rows. A DRAM row is a set of DRAM
cells that share a wire called the \emph{wordline}, which the \emph{local row
decoder} of the subarray drives after fully decoding the row address. In a
subarray, a column of cells shares a wire, referred to as the \emph{bitline},
that connects the column of cells to a \emph{sense amplifier}.  The sense
amplifier is the circuitry used to read and modify the data of a DRAM cell.
\hh{The row} of sense amplifiers in the subarray is referred to as the \emph{local
row-buffer}.  To access a DRAM cell, the corresponding DRAM row first needs to
be copied into the local row-buffer, which connects to the internal I/O bus via
the \emph{global row-buffer}.

Figure~\ref{subfig:dram_cell} illustrates a DRAM cell, which is composed of a
\emph{storage capacitor} and \emph{access transistor}. A DRAM cell stores a
single bit of information based on the charge level of the capacitor. The data
stored in the cell is interpreted as a ``1'' or ``0'' depending on whether the
charge stored in the cell is above or below a certain threshold. Unfortunately,
the capacitor and the access transistor are not ideal circuit components and
have \emph{charge leakage paths}. Thus, to ensure that the cell does not leak
charge to the point where the bit stored in the cell flips, the cell needs to
be periodically \emph{refreshed} to fully restore its original charge.

\subsubsection{DRAM Commands}
\label{subsubsec:dram_cmds}

~The memory controller issues a set of DRAM commands to access data in the DRAM
chip. To perform a read or write operation, the memory controller first needs
to \emph{open} a row, i.e., copy the data of the cells in the row to the
row-buffer. To open a row, the memory controller issues an \emph{activate
(ACT)} command to a bank by specifying the address of the row to open. The
memory controller can issue \emph{ACT} commands to different banks in
consecutive DRAM bus cycles to operate on \emph{multiple banks in parallel}.
After opening a row \hh{in a bank}, the memory controller issues either a \emph{READ} or a
\emph{WRITE} command to read or write a DRAM word (which is typically equal to
64 bytes) \hh{within the open row}. 
\hh{An open row can serve multiple \emph{READ} and \emph{WRITE} requests
without incurring precharge and activation delays.}
A DRAM row typically contains 4-8 KiBs of data. To access data from
another DRAM row \hh{in} the same bank, the memory controller must first close the
currently open row by issuing a \emph{precharge (PRE)} command. The memory
controller also periodically issues \emph{refresh (REF)} commands to prevent
data loss due to charge leakage.

\subsubsection{DRAM Cell Operation}
\label{subsubsec:dram_operation}


~We describe DRAM operation by explaining the steps involved in reading data
from \hh{a DRAM cell}.\footnote{Although we focus only on reading data, steps
involved in a write operation are similar.} The memory controller initiates
each step by issuing a DRAM command. Each step takes a certain amount of time
to complete, and thus, a DRAM command is typically associated with one or more
timing constraints known as \emph{timing parameters}. It is the responsibility
of the memory controller to satisfy these timing parameters in order to ensure
\emph{correct} DRAM operation.

\begin{figure}[h] \centering 
\includegraphics[width=\linewidth]{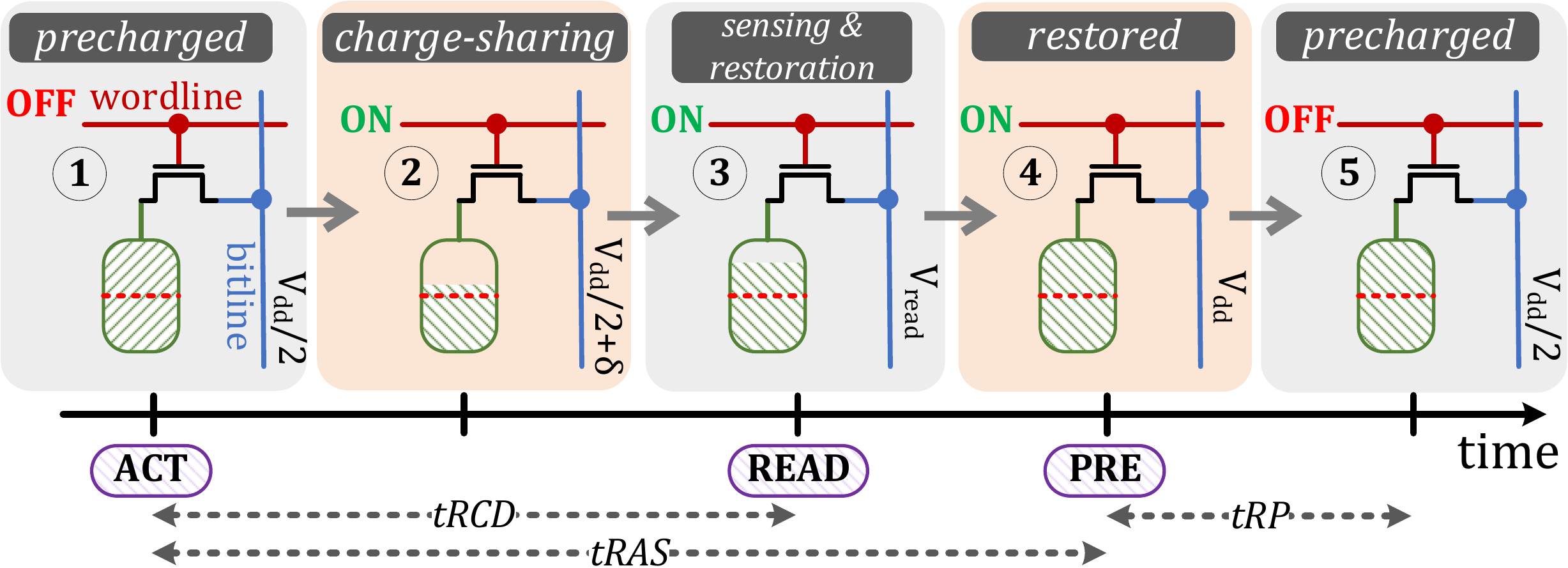} 
\caption{Command sequence for reading data from DRAM and the state of a DRAM cell during each \hhtwo{related} step.}  
\label{fig:dram_op}
\end{figure}

In Figure~\ref{fig:dram_op}, we show how the state of a DRAM cell changes
during the steps involved in a read operation. Each DRAM cell diagram
corresponds to the state of the cell at exactly the tick \hh{mark} on the time
axis. Each command (\jkfour{shown} in purple boxes below the \hhtwo{time} axis)
is issued by the memory controller at the \hh{corresponding} tick \hh{mark}.
Initially, the cell is in a \emph{precharged} state~\incircle{1}. When
precharged, the capacitor of the cell is disconnected from the bitline since
the wordline is not asserted and thus the access transistor is off. The bitline
voltage is stable at $\frac{V_{dd}}{2}$ and is ready to \hh{be perturbed}
towards the \hh{voltage level} of the cell \hh{capacitor} upon enabling the
access transistor. 

To read data from a cell, the memory controller first needs to perform
\emph{row activation} by issuing an \emph{ACT} command. During row activation
\hh{(\incircle{2}),} the row decoder asserts the wordline that connects the
storage capacitor of the cell to the bitline by enabling the access transistor.
At this point, the capacitor charge perturbs the bitline \hh{via the}
\emph{charge sharing} \hh{process}. Charge sharing continues until the
capacitor and bitline voltages reach an equal value of $\frac{V_{dd}}{2} +
\delta$. After charge sharing \hh{(\incircle{3}),} the sense amplifier begins
driving the bitline towards \hh{either} $V_{dd}$ or $0V$ depending on the
direction of the perturbation in the charge sharing step.  This step\hh{, which
amplifies} the voltage level \hh{on the bitline as well as the cell} is called
\emph{charge restoration}. Although charge restoration continues until the
original capacitor charge is fully replenished (\incircle{4}), the memory
controller can issue a \emph{READ} command to safely read data from the
activated row \hh{before the capacitor charge is fully replenished}. A
\emph{READ} command can \hasan{reliably} be issued when the bitline voltage
reaches \hh{the} voltage level $V_{read}$.  To ensure that the read occurs
after the bitline reaches $V_{read}$, the memory controller inserts a time
interval $t_{RCD}$ between the \emph{ACT} and \emph{READ} commands. It is the
responsibility of the DRAM \hh{manufacturer} to ensure that their DRAM chip
operates safely as long as the \hh{memory controller obeys the} $t_{RCD}$
timing parameter, which is defined in the DRAM standard~\cite{2014lpddr4}. If
the memory controller issues a \emph{READ} command before $t_{RCD}$ elapses,
the bitline voltage may be below $V_{read}$, which can lead to \hh{the} reading
\hh{of} a wrong value. 

To return a cell to its precharged state, the voltage in the cell must first be
fully restored. A cell is expected to be fully \hh{restored when the memory
controller satisfies} \hh{a} time \hh{interval dictated by} $t_{RAS}$
\hh{after} issuing the \emph{ACT} command. Failing to satisfy $t_{RAS}$ may
lead to \hh{insufficient amount of charge \hh{to be} restored} in the cells of
the accessed row. A subsequent activation of the row can then result in the
reading \hh{of} incorrect data from the cells.

Once the cell is successfully \emph{restored} \hh{(\incircle{4}),} the
memory controller can issue a \emph{PRE} command to close the
\hh{currently-open} row to prepare the bank for an access to another row.
The cell returns to the precharged state (\incircle{5}) after waiting for
\hh{the} timing parameter $t_{RP}$ following the \emph{PRE} command.
Violating $t_{RP}$ \hh{may prevent} the sense amplifiers from fully driving
the bitline back to $\frac{V_{dd}}{2}$, \hh{which may later result in the row
to be activated with too small amount of charge in its cells,} \hhtwo{potentially
preventing the sense amplifiers to read the data correctly}.

For correct DRAM operation, it is critical for the memory controller to ensure
that the DRAM timing parameters defined in the DRAM specification are
\emph{not} violated. Violation of the timing parameters may lead to
\hh{incorrect data to be read from the DRAM,} and \hh{thus cause} unexpected
program behavior~\cite{lee2015adaptive, khan2016parbor, hassan2017softmc,
chang2016understanding, chang2017understanding, chang2017thesis,
lee-sigmetrics2017}. In this work, we study \hhtwo{the} failure modes \hhtwo{due to
violating DRAM timing parameters} and explore their application \hh{to}
reliably generating true random numbers. 

\subsection{True Random Number Generators}
\label{subsec:trng}

A \emph{true random number generator (TRNG)} requires physical processes (e.g.,
radioactive decay, thermal noise, Poisson noise) to construct a bitstream of
random data. Unlike pseudo-random number generators, the random numbers
generated by a TRNG do \emph{not} depend on the \hh{previously-generated} numbers
and \emph{only} depend on the random noise obtained from physical processes.
TRNGs are usually validated using statistical tests such as
NIST~\cite{rukhin2001statistical} or DIEHARD~\cite{marsaglia2008marsaglia}. A
TRNG typically consists of 1) an \emph{entropy source}, 2) a \emph{randomness
extraction technique}, and sometimes 3) a \emph{post-processor}, which improves
the randomness of the extracted data often at the expense of throughput. These
three components are typically used to reliably generate true random
numbers\hh{~\cite{stipvcevic2014true, sunar2007provably}}.

\textbf{Entropy Source.} The entropy source is a critical component of a random
number generator, as its amount of entropy affects the unpredictability and the
throughput of the generated random data. Various physical phenomena can be
used as entropy sources. In the domain of electrical circuits, thermal and
Poisson noise, jitter, and circuit metastability have been proposed as processes
that have high entropy~\cite{wang2012flash, ray2018true, holcomb2007initial,
holcomb2009power, van2012efficient, mathew20122, brederlow2006low,
tokunaga2008true, bhargava2015robust}. To ensure robustness, the entropy source
should not be visible or modifiable by an adversary.  Failing to satisfy that
requirement would result in generating predictable data, and thus put the
system into a state susceptible to security attacks.

\textbf{Randomness Extraction Technique.} The randomness extraction
technique harvests random data from an entropy source. A good randomness
extraction technique should have two key properties. First, it should have
high throughput, i.e., extract as much as randomness possible in a short amount
of time~\cite{kocc2009cryptographic, stipvcevic2014true}\hh{, especially
important for applications that require high-throughput random number
generation (e.g., security applications~\cite{gutterman2006analysis, von2007dual, kim2017nano,
drutarovsky2007robust, kwok2006fpga, cherkaoui2013very, zhang2017high,
quintessence2015white, bagini1999design, rock2005pseudorandom, ma2016quantum,
stipvcevic2014true, barangi2016straintronics, tao2017tvl, botha2005gammaray,
mathew20122, yang201416}, scientific simulation~\cite{ma2016quantum,
botha2005gammaray})}. Second, it should not disturb the physical
process~\cite{kocc2009cryptographic, stipvcevic2014true}. Affecting the entropy
source during the randomness extraction process would make the harvested data
predictable, lowering the reliability of the TRNG.

\textbf{Post-processing.} Harvesting randomness from a physical phenomenon
\emph{may} produce bits that are biased or
correlated~\cite{kocc2009cryptographic, rahman2014ti}. In such a case, a
post-processing step, which is also known as \emph{de-biasing}, is applied to
eliminate the bias and correlation. The post-processing step also provides
protection against environmental changes and adversary
tampering~\cite{kocc2009cryptographic, rahman2014ti, stipvcevic2014true}.
\hhtwo{Well-known} post-processing techniques are the von Neumann
corrector~\cite{jun1999intel} and cryptographic hash functions such as
SHA-1~\cite{eastlake2001us} or MD5~\cite{rivest1992md5}. These post-processing
steps work well, but generally result in decreased throughput (e.g., up to
80\%~\cite{kwok2011comparison}).

\section{Motivation and Goal} 
\label{sec:motivation}

True random numbers sampled from physical phenomena have a number of real-world
applications from system security~\hh{\cite{bagini1999design,
rock2005pseudorandom, stipvcevic2014true}} to recreational
entertainment~\hh{\cite{stipvcevic2014true}}. As user data
privacy becomes a \emph{highly-sought} commodity in Internet-of-Things (IoT)
and mobile devices, enabling primitives that provide security on such systems
becomes \jk{critically important}~\cite{lin2017survey, ponemon2017sec, zhang2017high}.
Cryptography is one typical method for securing systems against various attacks
by encrypting the system's data with keys generated with true random values.
Many cryptographic algorithms require random values to generate keys in many
standard protocols (e.g., TLS/SSL/RSA/VPN keys) to either 1) encrypt network
packets, file systems, and data, 2) select internet protocol sequence numbers
(TCP), or 3) generate data padding values~\cite{gutterman2006analysis,
von2007dual, kim2017nano, drutarovsky2007robust, kwok2006fpga,
cherkaoui2013very, zhang2017high, quintessence2015white}. TRNGs are also
commonly used in authentication protocols and in countermeasures against
hardware attacks~\cite{cherkaoui2013very}, in which psuedo-random number
generators (PRNGs) \hh{are shown to be insecure}~\cite{von2007dual,
cherkaoui2013very}. To keep up with the \emph{ever-increasing} rate of
secure data creation, especially with the growing number of commodity
data-harvesting devices (e.g., IoT and mobile devices), the ability to generate
true random numbers with \emph{high throughput and low latency} becomes ever
more relevant to maintain user data privacy. In addition,
\emph{high-throughput} TRNGs are \hh{already} \emph{essential} components of
various important applications such as scientific
simulation~\hh{\cite{ma2016quantum, botha2005gammaray}}, industrial testing,
statistical sampling, randomized algorithms, and recreational
entertainment~\cite{bagini1999design, rock2005pseudorandom, ma2016quantum,
stipvcevic2014true, barangi2016straintronics, tao2017tvl, botha2005gammaray,
zhang2017high, mathew20122, yang201416}. 

A \emph{\hh{widely-available,} high-throughput, low-latency} TRNG will enable
all previously mentioned applications that rely on TRNGs, including improved
\jktwo{security and privacy} in most systems that are known to be vulnerable to
attacks~\cite{lin2017survey, ponemon2017sec, zhang2017high}, as well as enable
research that we may not anticipate at the moment. One such direction is using
a one-time pad \hhthree{(i.e., a private key used to encode and decode only a single message)} with quantum key
distribution, which requires at least \emph{4Gb/s} \hh{of true random number
generation throughput}~\cite{wang2016theory, clarke2011robust, lu2015fpga}.
Many \emph{high-throughput} TRNGs have been recently
proposed~\cite{kwok2006fpga, tsoi2007high, zhang201568, cherkaoui2013very,
barangi2016straintronics, zhang2017high, ning2015design, quintessence2015white,
wang2016theory, mathew20122, yang201416, bae20173, fischer2004high,
gyorfi2009high}, and the availability of these high-throughput TRNGs can enable
\hh{a wide range of new applications with improved security} \jktwo{and privacy}.

DRAM offers a promising substrate for developing an effective and
widely-available TRNG due to the prevalence of DRAM throughout all modern
computing systems ranging from microcontrollers to supercomputers. A
high-throughput DRAM-based TRNG would help enable widespread adoption of
applications that are today limited to only select architectures equipped with
dedicated high-performance TRNG engines. Examples of such applications include
high-performance scientific simulations and cryptographic applications for
securing devices and communication protocols, both of which would run
\hhthree{much} more
\jk{efficiently on \hhthree{mobile devices, embedded devices,} or microcontrollers with the availability
of higher-throughput TRNGs \hhthree{in} the system.} 



In terms of the CPU architecture itself, a high-throughput DRAM-based TRNG
\jkfour{could} help the memory controller to improve scheduling
decisions~\cite{usui2016dash, mutlu2009parallelism, ausavarungnirun2012staged,
subramanian2016bliss, kim2010thread, mutlu2007stall, subramanian2013mise,
subramanian2015application} and \jkfour{enable the implementation} a
truly-randomized version of PARA~\cite{kim2014flipping} (i.e., a protection
mechanism against the RowHammer \jktwo{vulnerability~\cite{kim2014flipping,
mutlu2017rowhammer}}).  \jktwo{Furthermore, a DRAM-based TRNG would likely have
additional hardware and software applications as system designs become more
capable and increasingly security-critical.} 

In addition to traditional computing paradigms, DRAM-based TRNGs can
benefit processing-in-memory (PIM) architectures~\cite{ghose2018enabling,
seshadri2017simple}, which co-locate logic \hhthree{within or near} memory \jktwo{to overcome the
large bandwidth and energy bottleneck caused by} the memory bus and leverage
the \emph{significant} data parallelism available within the DRAM chip itself.
Many prior works provide primitives for PIM or exploit PIM-enabled systems for
workload acceleration~\cite{ahn2015scalable, ahn2015pim, lee2015simultaneous,
seshadrifast, seshadri2013rowclone, seshadri2015gather, seshadri2017ambit,
liu2017concurrent, seshadri2017simple, pattnaik2016scheduling,
babarinsa2015jafar, farmahini2015nda, gao2015practical, gao2016hrl,
hassan2015near, hsieh2016transparent, morad2015gp, sura2015data, zhang2014top,
hsieh2016accelerating, boroumand2017lazypim, chang2016low, kim2018grim,
ghose2018enabling, boroumand2018google, seshadri2016simple}. A low-latency,
high-throughput DRAM-based TRNG can enable PIM applications to source random
values \emph{directly within the memory itself}, thereby enhancing the overall
potential, \jktwo{security, and privacy,} of PIM-enabled architectures. For
example, \hhf{in applications that require true random numbers, a DRAM-based
TRNG can enable large contiguous code segments to execute in memory, which
would reduce communication with the CPU, and thus improve system efficiency. A
DRAM-based TRNG can also enable security tasks to run completely in memory.
This would remove the dependence of PIM-based security tasks on an I/O channel
and would increase \jkfive{overall} system security.}

We posit, based on analysis done in prior works~\cite{kocc2009cryptographic,
jun1999intel, schindler2002evaluation}, that an \emph{effective} TRNG must
satisfy \emph{six} key properties: \jktwo{it} must 1) have \hh{low} implementation
cost, 2) be fully non-deterministic such that it is impossible to predict the
next output given complete information about how the mechanism operates, 3)
provide a continuous stream of true random numbers with high throughput, 4)
provide true random numbers with low latency, 5) exhibit low system
interference, i.e., not significantly \hh{slow down} concurrently-running
applications, and 6) generate random values with low energy overhead.

To this end, our \textbf{goal} in this work, is to provide a
\hh{widely-available} TRNG for DRAM
devices that satisfies all six key properties of an effective TRNG. 


\section{Testing Environment}
\label{sec:methodology} 

In order to test our hypothesis that DRAM cells are an effective source of
entropy when accessed with reduced DRAM timing parameters, we developed an
infrastructure to characterize modern LPDDR4 DRAM chips.
We also use an infrastructure for DDR3 DRAM chips,
SoftMC~\cite{hassan2017softmc, softmc-safarigithub}, to demonstrate
empirically that our proposal is applicable beyond the LPDDR4
technology. Both testing environments give us precise control over DRAM
commands and DRAM timing parameters as verified with a logic analyzer probing
the command bus.

We perform all tests, unless otherwise specified, using a total of 282 2y-nm
LPDDR4 DRAM chips from three major manufacturers in a thermally-controlled
chamber held at 45$^{\circ}$C. For consistency across results, we
precisely stabilize the ambient temperature using heaters and fans
controlled via a microcontroller-based proportional-integral-derivative (PID)
loop to within an accuracy of 0.25$^{\circ}$C and a reliable range of
40$^{\circ}$C to 55$^{\circ}$C.  We maintain DRAM temperature at 15$^{\circ}$C
above ambient temperature using a separate local heating source. We use
temperature sensors to smooth out temperature variations caused by self-induced
heating.

We also use a separate infrastructure, based on open-source
SoftMC~\cite{hassan2017softmc, softmc-safarigithub}, to validate our mechanism
on 4 DDR3 DRAM chips from a single manufacturer. SoftMC enables precise control
over timing parameters, and we house the DRAM chips inside \jktwo{another}
temperature chamber to maintain a stable ambient testing temperature (with the
same temperature range as the temperature chamber \jktwo{used for the LPDDR4
devices}).  

To explore the various effects of temperature, short-term aging, and
circuit-level interference (in Section~\ref{sec:dlrng_characterization}) on
activation failures, we reduce the $t_{RCD}$ parameter from the default
$18ns$ to $10ns$ for all experiments, unless otherwise stated.
Algorithm~\ref{alg:testing_lat} explains the general testing methodology we
use to induce activation failures. First, we write a data pattern to the
    \begin{algorithm}[tbh]\footnotesize
        \SetAlgoNlRelativeSize{0.7}
        \SetAlgoNoLine
        \DontPrintSemicolon
        \SetAlCapHSkip{0pt}
        \caption{DRAM Activation Failure Testing}
        \label{alg:testing_lat} 

        \textbf{DRAM\_ACT\_failure\_testing($data\_pattern$, $DRAM\_region$):} \par 
        ~~~~write $data\_pattern$ (e.g., solid 1s) into all cells in $DRAM\_region$ \par 
        ~~~~set low $t_{RCD}$ for ranks containing $DRAM\_region$ \par 
        ~~~~\textbf{foreach} $col$ in $DRAM\_region$: \par 
        ~~~~~~~~\textbf{foreach} $row$ in $DRAM\_region$: \par 
        ~~~~~~~~~~~~$activate(row)$ \textcolor{gray}{~~~~// fully refresh cells } \par 
        ~~~~~~~~~~~~$precharge(row)$ \textcolor{gray}{~// ensure next access activates the row} \par
        ~~~~~~~~~~~~$activate(row)$ \par 
        ~~~~~~~~~~~~$read(col)$ \textcolor{gray}{~~~~~~~~~~~// induce activation failure on col} \par
        ~~~~~~~~~~~~$precharge(row)$ \par
        ~~~~~~~~~~~~record activation failures to storage \par 
        ~~~~set default $t_{RCD}$ for DRAM ranks containing $DRAM\_region$ \par 
    \end{algorithm}
region of DRAM under test (Line~2). Next, we reduce the $t_{RCD}$
parameter to begin inducing activation failures (Line~3). We then access
the DRAM region in column order (Lines~4-5) in order to ensure that each DRAM
access is to a closed DRAM row and thus requires an activation. This enables
each access to induce activation failures in DRAM. Prior to each
reduced-latency read, we first refresh the target row such that each cell has
the same amount of charge each time it is accessed with a reduced-latency read.
We effectively refresh a row by issuing an activate (Line~6) followed by a
precharge (Line~7) to that row. We then induce the activation failures by
issuing consecutive activate (Line~8), read (Line~9), and precharge (Line~10)
commands. Afterwards, we record any activation failures that we observe
(Line~11).  We find that this methodology enables us to quickly induce
activation failures across \emph{all} of DRAM, and \jktwo{minimizes} testing
time. 

\section{Activation Failure Characterization} 
\label{sec:dlrng_characterization} 

To demonstrate the viability of using DRAM cells as an entropy source for
random data, we explore and characterize DRAM failures when employing a
reduced DRAM activation latency ($t_{RCD}$) across 282 LPDDR4 DRAM chips. We
also compare our findings against \mpx{those of} prior works that study an
older generation of DDR3 DRAM chips~\cite{chang2016understanding,
lee-sigmetrics2017, lee2015adaptive, kim2018solar} to \jk{cross-}validate our
infrastructure. \jk{To understand the effects of changing environmental
conditions on a DRAM cell that is used as a source of entropy, we rigorously
characterize DRAM cell behavior as we vary four environmental conditions.
First, we study the effects of DRAM array design-induced variation (i.e., the
spatial distribution of activation failures in DRAM). Second, we study data
pattern dependence (DPD) effects on DRAM cells. Third, we study the effects of
temperature variation on DRAM cells. Fourth, we study a DRAM cell's activation
failure probability over time.} We present several key observations that
support the viability of a mechanism that generates random numbers by
accessing DRAM cells with a reduced $t_{RCD}$.  In
Section~\ref{section:DLRNG_mechanism}, we discuss a mechanism to effectively
\jk{sample DRAM cells to extract true random numbers while minimizing the effects
of environmental condition variation (presented in this section) on the DRAM
cells.} 

\subsection{Spatial Distribution of Activation Failures}
\label{subsec:spatial_char} 

\begin{figure}[h] 
    \centering \includegraphics[width=0.7\linewidth]{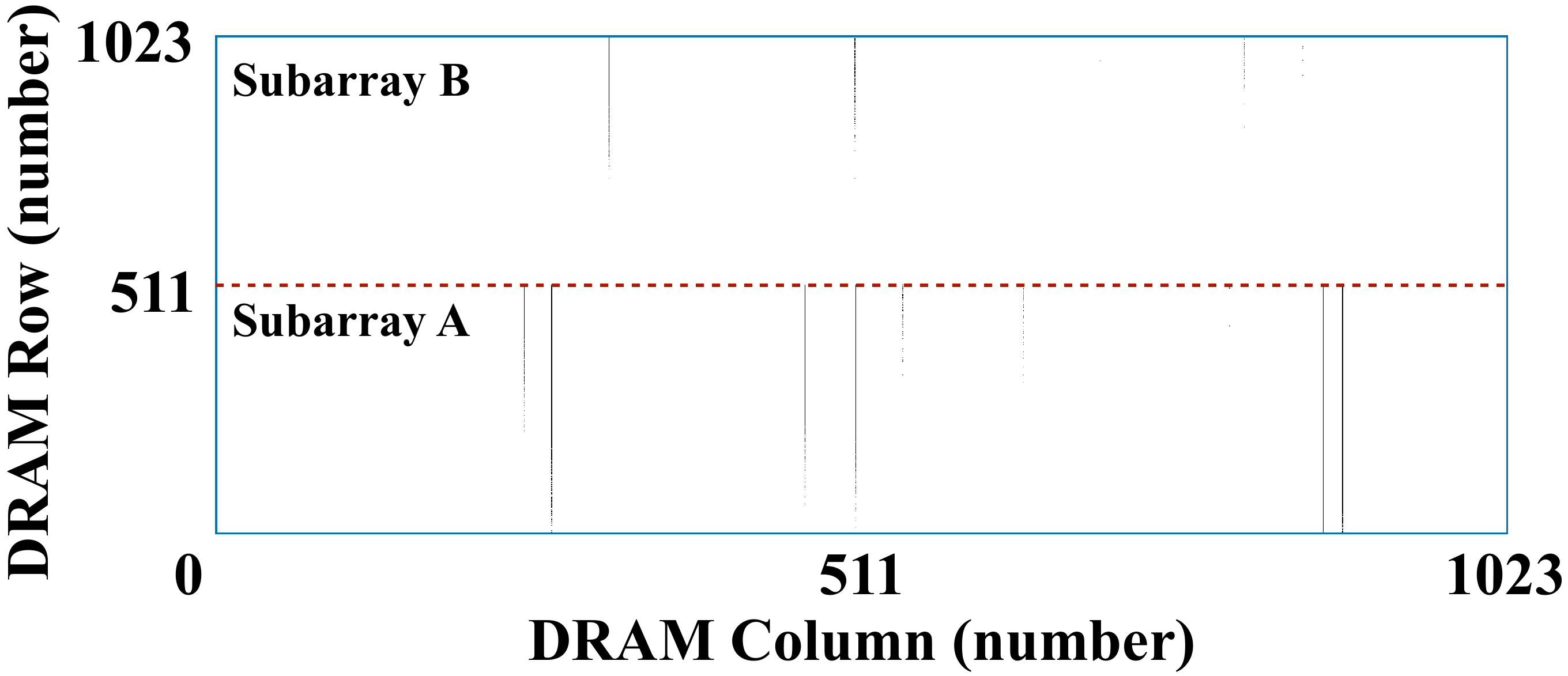} 
    \caption{Activation failure bitmap in $1024\times1024$ cell array.}
    \label{fig:spatial_failures} 
\end{figure}

To study \mpx{which} regions of DRAM \mpx{are better suited to} generating
random data, we first visually inspect the spatial distributions of activation
failures \mpx{both} across DRAM chips and within each chip \mpx{individually}. 
Figure~\ref{fig:spatial_failures} plots the spatial distribution of activation
failures in a \emph{representative} $1024\times1024$ array of DRAM cells
\mpx{taken} from a single DRAM chip\mpx{. Every} observed activation
failure is marked in black. We make two observations. First, we observe that
each contiguous region of 512 DRAM rows\footnote{\jk{We note that subarrays
have either 512 or 1024 (not shown) rows depending on the manufacturer of the
DRAM device.}} \mpx{consists} of \mpx{repeating} rows with the same set (or
subset) of \jk{column bits} that are prone to activation failures. As shown in
the figure, rows 0 to 511 have the same 8 (or a subset of the 8) \jk{column
bits} failing in the row, and rows 512 to 1023 have the same 4 (or a subset of
the 4) \jk{column bits} failing in the row. We hypothesize that these
contiguous regions reveal the DRAM subarray architecture \mpx{as a result of
variation across} the local sense amplifiers in the subarray. \jkthree{We
indicate the two subarrays in Figure~\ref{fig:spatial_failures} as Subarray A
and Subarray B.} A ``weaker'' local sense amplifier results in cells that share
\mpx{its} respective \emph{local bitline} in the subarray \mpx{having} an
increased probability of failure. For this reason, we observe \mpx{that}
activation failures \mpx{are} localized to a few columns within a DRAM subarray
as shown in Figure~\ref{fig:spatial_failures}. Second, we observe that within a
subarray, the activation failure probability \mpx{increases} across rows (i.e.,
activation failures are \emph{more} likely to occur in \jk{higher-numbered}
rows in the subarray and \mpx{are} \emph{less} likely in \jk{lower-numbered}
rows in the subarray). This can be seen \jkthree{from} the fact that more cells
fail in \jk{higher-numbered} rows in the subarray (i.e., there are more black
marks higher in each subarray). We hypothesize that the \jkthree{failure
probability of a cell attached to a local bitline} correlates \mpx{with} the
distance between the row and the local sense amplifiers, and further rows have
less time to amplify their data due to the signal propagation delay in a
bitline. These observations are similar to \mpx{those made in} prior
studies~\cite{lee-sigmetrics2017, chang2016understanding, lee2015adaptive,
kim2018solar} on DDR3 devices.


We next study the granularity at which \mpx{we can induce} activation failures
when accessing a row. We observe (not shown) that activation failures
\mpx{occur} \emph{only} within the \mpx{first} cache line that is \mpx{accessed}
 immediately following an activation. No subsequent access to an
\jk{already} \emph{open} row results in activation failures. This is because
cells within the \emph{same} row have a longer time to restore their cell
charge (Figure~\ref{fig:dram_op}) when they are accessed after the row
\mpx{has already been opened}.  We draw two key conclusions: 1) the region
\emph{and} bitline of DRAM being accessed affect the number of observable
activation failures, and 2) different \mpx{DRAM} subarrays \emph{and}
different local bitlines exhibit varying levels of entropy.

\subsection{Data Pattern Dependence} 
\label{subsec:dpd} 

To understand the data pattern dependence of activation failures and DRAM cell
entropy, we study \mpy{how effectively we can discover failures using
different data patterns across multiple rounds of testing.} Our \emph{goal} in
this \mpy{experiment} is to determine \mpy{which} data pattern results in the
highest entropy such that we can generate random values with high throughput.
Similar to prior works~\cite{patel2017reaper, liu2013experimental} that
extensively describe the data patterns, we analyze a total of 40 unique data
patterns: solid 1s, checkered, row stripe, column stripe, 16 walking 1s,
\emph{and} the inverses of all 20 aforementioned data patterns.

\begin{figure}[h] 
    \centering \includegraphics[width=\linewidth]{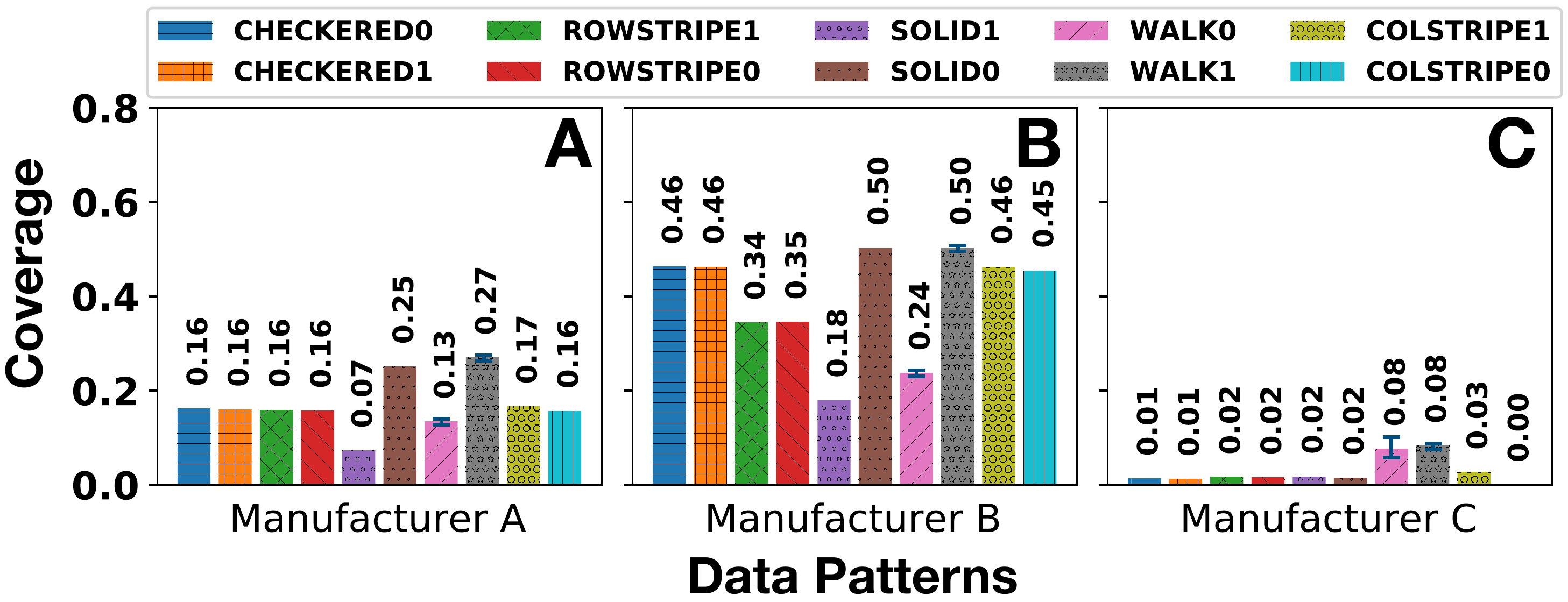} 
    \caption{Data pattern dependence of DRAM cells prone to activation failure over 100 iterations} 
    \label{fig:dpd_coverage}
\end{figure}

Figure~\ref{fig:dpd_coverage} plots the \mpy{ratio of activation failures
discovered by a particular data pattern after 100 iterations of
Algorithm~\ref{alg:testing_lat} relative to the \emph{total} number of
failures discovered by \emph{all} patterns for a representative chip from each
manufacturer}. \mpy{We call this metric \emph{coverage} because it
indicates the effectiveness of a single data pattern to identify all possible
DRAM cells that are prone to activation failure. We show results for each
pattern individually except for the WALK1 and WALK0 patterns, for which we
show the mean (bar) and minimum/maximum (error bars) coverage across all 16
iterations of each walking pattern.}

\mpy{We make three key observations from this experiment.} \mpy{First}, we
find that testing with different data patterns identifies different subsets of
the total set of possible activation failures.  This indicates that \jk{1)
different data patterns cause different DRAM cells to fail and 2) specific
data patterns induce more activation failures than others. Thus,} certain data
patterns may extract more entropy from a DRAM cell array than other data
patterns. \mpy{Second}, we find that, of \emph{all} 40 tested data patterns,
each of the 16 \emph{walking 1s}, for a given device, provides a \jk{similarly
high} coverage, regardless of the manufacturer. This high coverage is
similarly provided by only one other data pattern per manufacturer: solid 0s
for manufacturers A and B, and walking 0s for manufacturer C. \mpy{Third, if
we repeat this experiment (i.e., Figure~\ref{fig:dpd_coverage}) while varying
the number of iterations of Algorithm~\ref{alg:testing_lat}, the \emph{total
failure count} across all data patterns \emph{increases} as we increase the
number of iterations of Algorithm~\ref{alg:testing_lat}.} This indicates that
\mpy{not all DRAM cells fail deterministically when accessed with a reduced
$t_{RCD}$, providing a potential source of entropy for random number
generation.}

We next analyze \mpy{each cell's probability of failing when accessed with a
reduced $t_{RCD}$ (i.e., its  \emph{activation failure probability}) to
determine which data pattern most effectively identifies cells that provide
high entropy}. We note that DRAM cells with an activation failure probability
$F_{prob}$ of 50\% provide high entropy when accessed many times. With the
same data used to produce Figure~\ref{fig:dpd_coverage}, we study the
different data patterns with regard to the number of cells they cause to fail
50\% of the time. Interestingly, we find that the data pattern that
\jkfour{induces} the most failures overall does not necessarily find the most
number of cells that fail 50\% of the time. In fact, when searching for cells
with an $F_{prob}$ between 40\% and 60\%, \jk{we observe that} the data
\mpy{patterns that find} the \jk{highest number of cells} \mpy{are} solid 0s,
checkered 0s, and solid 0s for manufacturers A, B, and C, respectively.  We
conclude that: 1) due to manufacturing and design variation across DRAM
devices from different manufacturers, different data patterns \mpy{result in}
different failure probabilities in our DRAM devices, and 2) to provide high
entropy \mpy{when} accessing DRAM cells with a reduced $t_{RCD}$, we should
use the respective data pattern that finds the most number of cells with an
$F_{prob}$ of 50\% for DRAM devices from a given manufacturer.

Unless otherwise stated, in the rest of this paper, we use \mpy{the} solid 0s,
checkered 0s, and solid 0s data patterns for manufacturers A, B, and C,
respectively, to analyze $F_{prob}$ at the granularity of a single cell
\mpy{and} to study the effects of temperature and time on our sources of
entropy.

\subsection{Temperature Effects} 
\label{subsec:temp_effects}

\begin{figure}[h] 
    \centering \includegraphics[width=\linewidth]{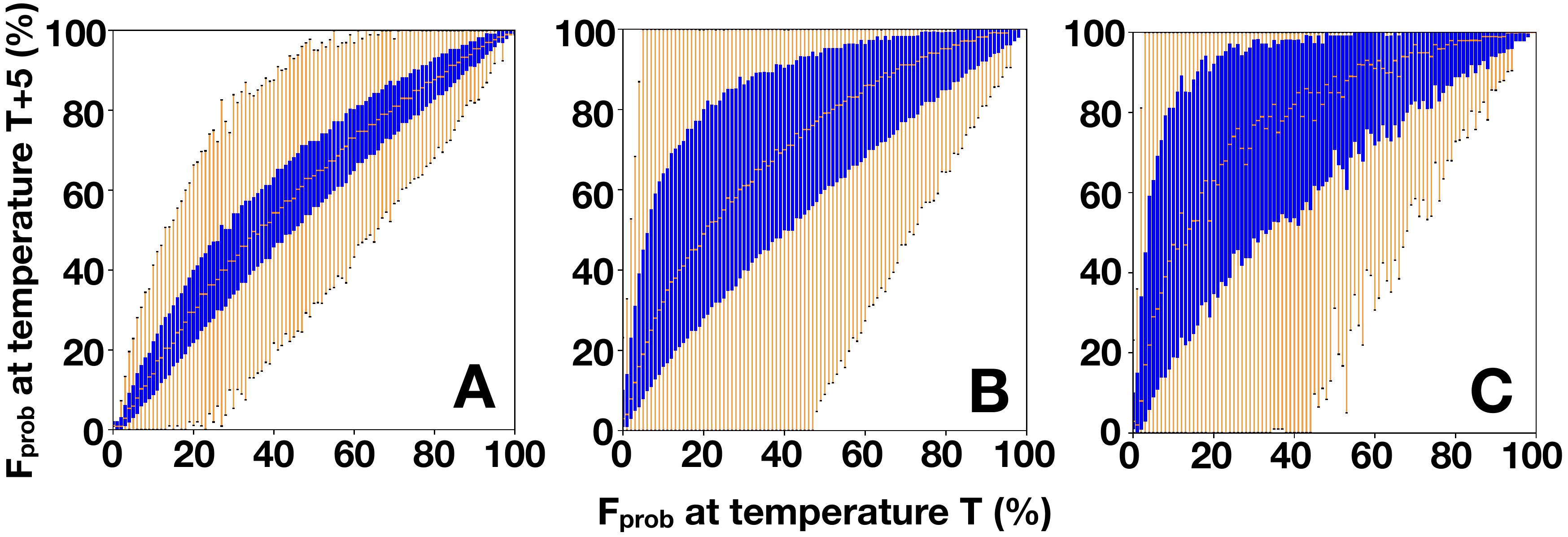} 
    \caption{Effect of temperature variation on failure \hhf{probability}} 
    \label{fig:temperature_effects} 
\end{figure}

In this section, we study whether temperature \mpy{fluctuations affect} a DRAM
cell's activation failure probability and thus the entropy that can be
extracted from the DRAM cell. To analyze \mpy{temperature effects}, we record
the $F_{prob}$ of cells \mpy{throughout} our DRAM devices \mpy{across 100
iterations of Algorithm~\ref{alg:testing_lat} at 5$^{\circ}$C increments}
between 55$^{\circ}$C and 70$^{\circ}$). Figure~\ref{fig:temperature_effects}
aggregates \mpy{results} across 30 DRAM modules from each DRAM manufacturer.
\jk{Each point in the figure represents how the $F_{prob}$ of a DRAM cell
changes as the temperature changes (i.e., ${\Delta}F_{prob}$). The x-axis
shows the $F_{prob}$ of a single cell at temperature $T$ (i.e., the baseline
temperature), and the y-axis shows the $F_{prob}$ of the same cell at
temperature $T+5$ (i.e., 5$^{\circ}$C above the baseline temperature).}
Because we test each cell at each temperature across 100 iterations, the
granularity \mpy{of} $F_{prob}$ on both the x- and y-axes is 1\%. For a given
$F_{prob}$ at temperature $T$ \jk{(x\% on the x-axis)}, we aggregate
\emph{all} respective $F_{prob}$ points at temperature $T+5$ \jk{(y\% on the
y-axis)} with box-and-whiskers plots\footnote{A box-and-whiskers plot
emphasizes the important metrics of a dataset's distribution. The box is
lower-bounded by the first quartile (i.e., the median of the first half of the
ordered set of data points) and upper-bounded by the third quartile (i.e., the
median of the second half of the ordered set of data points). The median falls
within the box. The \emph{inter-quartile range} (IQR) is the distance between
the first and third quartiles (i.e., box size).  Whiskers extend an additional
$1.5 \times IQR$ on either sides of the box.  We indicate outliers, or data
points outside of the range of the whiskers, with pluses.} to show how the
given $F_{prob}$ is affected by the increased DRAM temperature.  The
\emph{box} is drawn in blue and contains the \emph{median} drawn in red.  The
\emph{whiskers} are drawn in gray, and the \emph{outliers} are indicated with
orange pluses. 

We observe that $F_{prob}$ at temperature $T+5$ tends to be higher than
$F_{prob}$ at temperature $T$\jkfour{,} \mpy{as shown by} the blue region of the figure
(i.e., the boxes of the box-and-whiskers plots) \mpy{lying} above the $x~=~y$
line. However, \mpy{fewer} than 25\% of all data points fall below the $x~=~y$
line, indicating that a portion of cells have a lower $F_{prob}$ as
temperature is increased.

We observe that DRAM devices from different manufacturers are affected by
temperature differently. DRAM cells of manufacturer A have the \emph{least}
variation of ${\Delta}F_{prob}$ when temperature is increased since the boxes
of the box-and-whiskers plots are strongly correlated with the $x~=~y$ line.
\jk{How a DRAM cell's activation failure probability changes in DRAM devices
from \emph{other} manufacturers is unfortunately \emph{less} predictable
under} temperature change (i.e., a DRAM cell from manufacturers B or C has
higher variation in $F_{prob}$ change), but \mpy{the data still shows a strong
positive correlation between temperature and $F_{prob}$}.  We conclude that
temperature affects \jk{cell failure probability ($F_{prob}$)} to different
degrees depending on the manufacturer of the DRAM device, \mpy{but increasing
temperature generally increases the activation failure probability.}

\subsection{Entropy Variation over Time}
\label{subsec:short_term_variation}

To determine whether the failure probability of a DRAM cell \jk{changes over
time}, we complete 250 \emph{rounds} of recording the activation failure
probability of DRAM cells over the span of 15 days.  Each round consists of
accessing every cell in DRAM 100 times with a reduced $t_{RCD}$ value and
recording the failure probability for each individual cell (out of 100
iterations). We \mpy{find} that a DRAM cell's activation failure probability
does \emph{not} change significantly over time. This means that, \mpy{once we
identify} a DRAM cell that exhibits high entropy, we can rely on the cell to
maintain its high entropy over time.  We hypothesize that this is because 
\jkfour{a DRAM cell fails with high entropy} when process manufacturing
variation in peripheral and DRAM cell circuit elements combine such that, when
we read the cell using a reduced $t_{RCD}$ value, we induce a metastable state
resulting from the cell voltage falling between the reliable sensing margins
(i.e., falling close to $\frac{V_{dd}}{2}$)~\cite{chang2016understanding}.
Since manufacturing variation is fully determined at manufacturing time, a
\jkfour{DRAM cell's activation failure probability} is stable over time given
the same experimental conditions. In Section~\ref{subsec:cell_selection}, we
discuss our \mpy{methodology} for selecting DRAM cells for extracting stable
entropy, such that we can preemptively avoid longer-term aging effects that we
do not study in this paper.

\delete{\subsection{Effects of Reduced $t_{RCD}$ on DRAM Writes}} 
\delete{We rely on inducing activation failures on DRAM cells having the same data
pattern each time. This means that between every reduced $t_{RCD}$ access,
we must restore, i.e., rewrite, the data pattern in the recently accessed DRAM
word.  Therefore, we would like to study the effects that a reduced $t_{RCD}$
has on DRAM write accesses to see whether we need to reset $t_{RCD}$ to its
default value before restoring the data pattern in the DRAM cells. To test
whether DRAM writes are affected by a reduced $t_{RCD}$ value, we run two
experiments with our set of DRAM devices. First, we sweep the value of
$t_{RCD}$ between $10ns$ and $18ns$ and write a known data pattern across DRAM. We
then subsequently check the DRAM array for unexpected data and we observe no
failures in this range of $t_{RCD}$. } 
\delete{In our second experiment, we would like to verify that a shortened write
command does not result in side-effects such as lower voltage levels in the
written DRAM cells. Thus, for a range of $t_{RCD}$ values, we run the following
test.  For a $t_{RCD}$ value, we write a known data pattern into the DRAM
array, and disable DRAM refresh, so the charge in every cell will drain over
time. We then compare the failure rates across the DRAM array for a set
interval with refresh disabled. We note highly similar failure rates regardless
of the value of $t_{RCD}$ and conclude that writing to DRAM with a
significantly reduced $t_{RCD}$ has no impact on the quality of the writes.
This enables DRAM to service application writes between reduced $t_{RCD}$
accesses for true random number generation to minimize system interference.} 


\section{\mechanism: A DRAM-based TRNG} 
\label{section:DLRNG_mechanism} 

\jkfour{Based on} our rigorous analysis of \mpy{DRAM} activation failures
(presented in Section~\ref{sec:dlrng_characterization}), we propose \mechanism,
a flexible \mpy{mechanism} that provides high-throughput DRAM-based true random
number \mpy{generation} (TRNG) by sourcing entropy from a subset of DRAM cells
\jk{and is built fully within the memory controller.} \mechanism~is \mpy{based
on the \textbf{key observation}} that DRAM cells fail \mpy{probabilistically}
when accessed with reduced DRAM timing parameters\jk{, and this probabilistic
failure mechanism can be used as a source of true random numbers}.  While there
are many other timing parameters that we could \mpy{reduce} to induce failures
in DRAM~\cite{chang2016understanding, lee-sigmetrics2017, lee2015adaptive,
kim2018solar, lee2016reducing, chang2017thesis}, we focus specifically on
reducing $t_{RCD}$ below manufacturer-recommended values to study the resulting
activation failures.\footnote{We believe that reducing other timing parameters
could be used to generate true random values, but we leave their exploration to
future work.}

Activation failures occur as a result of reading the value from a DRAM cell
\emph{too soon} after sense amplification. This results in reading the value
at the sense amplifiers before the bitline voltage is amplified to an
\jk{I/O-readable} voltage level. The probability of reading incorrect data
from the DRAM cell therefore depends largely \jk{on the bitline's voltage} at
the time of reading the sense amplifiers. Because there is significant process
variation across the DRAM cells and I/O
circuitry~\cite{chang2016understanding, lee-sigmetrics2017, lee2015adaptive,
kim2018solar}, we observe a wide \mpy{variety} of failure probabilities for
different DRAM cells (as discussed in
Section~\ref{sec:dlrng_characterization}) for a given $t_{RCD}$ value\mpy{,
ranging from 0\% probability to 100\% probability}.

\textbf{We discover that a subset of cells fail at \mpy{$\jkfour{\sim}$}50\%
probability, and a subset of these cells fail randomly with high entropy
(shown in Section~\ref{subsec:diff_chips}).} In this section, we first discuss
our method of identifying such cells, \jk{which} we refer to as \emph{RNG
cells} (in Section~\ref{subsec:cell_selection}). Second, we describe the
mechanism with which D-RaNGe \emph{samples} RNG cells to extract random data
(Section~\ref{subsec:sampling_rng_cells}). Finally, we discuss a potential
design for integrating D-RaNGe \jk{in} a full system
(Section~\ref{subsec:full_system_integration}).

\subsection{RNG Cell Identification} 
\label{subsec:cell_selection} 

Prior to generating random data, we must first identify cells that are capable
of producing truly random output (i.e., RNG cells).  Our process of
identifying RNG cells involves reading every cell in the DRAM array 1000 times
with a \emph{reduced} $t_{RCD}$ and approximating each cell's Shannon
entropy~\cite{shannon1948mathematical} by counting the occurrences of 3-bit
symbols across its 1000-bit stream. We identify cells that \mpy{generate an
approximately equal} number of every \mpy{possible} 3-bit symbol ($\pm10\%$ of
the number of expected symbols) as RNG cells. 

We find that \mpy{RNG cells provide unbiased output, meaning that a}
post-processing step (described in Section~\ref{subsec:trng}) is \emph{not}
necessary \mpy{to provide sufficiently high entropy for random number
generation}. We also find that RNG cells \emph{maintain high entropy across
system reboots}. In order to account for our observation that entropy from an
RNG cell changes depending on the DRAM temperature
\jk{(Section~\ref{subsec:temp_effects})}, we identify reliable RNG cells at
each temperature and store their locations in the memory controller. Depending
on the DRAM temperature \jk{at the time an application requests random
values,} D-RaNGe samples the appropriate RNG cells. To ensure that DRAM aging
does not negatively impact the reliability of RNG cells, we require
re-identifying the set of RNG cells at regular intervals. From our observation
that entropy does not change significantly over a tested 15 day period of
sampling RNG cells
\jk{(Section~\ref{subsec:short_term_variation})}, we expect the interval of
re-identifying RNG cells to be at least 15 days long.  Our RNG cell
identification process is effective at identifying cells that are reliable
entropy sources for random number generation, and we quantify their randomness
using the NIST test suite for randomness~\cite{rukhin2001statistical} in
Section~\ref{subsec:nist}. 

\subsection{Sampling RNG Cells for Random Data} 
\label{subsec:sampling_rng_cells} 

Given the availability of these RNG cells, we use our observations in
Section~\ref{sec:dlrng_characterization} to design a high-throughput TRNG that
quickly and repeatedly samples RNG cells with reduced DRAM timing parameters.
Algorithm~\ref{alg:lat_rng} demonstrates the key components of \mechanism~that
enable us to generate random numbers with high throughput. 
    \begin{algorithm}[tbh]\footnotesize
        \SetAlgoNlRelativeSize{0.7}
        \SetAlgoNoLine
        \DontPrintSemicolon
        \SetAlCapHSkip{0pt}
        \caption{\mechanism:~A DRAM-based TRNG} 
        \label{alg:lat_rng}

        \textbf{\mechanism($num\_bits$):} \textcolor{gray}{~~~// \emph{num\_bits}: number of random bits requested} \par 
        ~~~~$DP$: a known data pattern that results in high entropy \par 
        ~~~~select 2 DRAM words with RNG cells in distinct rows in each bank \par
        ~~~~write $DP$ to chosen DRAM words and their neighboring cells \par
        ~~~~get exclusive access to rows of chosen DRAM words and nearby cells \par 
        ~~~~set low $t_{RCD}$ for DRAM ranks containing chosen DRAM words \par 
        ~~~~\textbf{for} each bank: \par 
        ~~~~~~~~read data in $DW_1$ \textcolor{gray}{~~// induce activation failure} \par 
        ~~~~~~~~write the read value of $DW_1$'s RNG cells to $bitstream$ \par
        ~~~~~~~~write original data value back into $DW_1$ \par 
        ~~~~~~~~memory barrier \textcolor{gray}{~~~~// ensure completion of write to $DW_1$} \par 
        ~~~~~~~~read data in $DW_2$ \textcolor{gray}{~// induce activation failure} \par 
        ~~~~~~~~write the read value of $DW_2$'s RNG cells to $bitstream$ \par 
        ~~~~~~~~write original data value back into $DW_2$ \par 
        ~~~~~~~~memory barrier \textcolor{gray}{~~~// ensure completion of write to $DW_2$} \par 
        ~~~~~~~~\textbf{if} $bitstream_{size}$$~\geq~num\_bits$: \par 
        ~~~~~~~~~~~~break \par 
        ~~~~set default $t_{RCD}$ for DRAM ranks of the chosen DRAM words \par 
        ~~~~release exclusive access to rows of chosen words and nearby cells \par 
    \end{algorithm}
\mechanism~ takes in $num\_bits$ as an argument, which is defined as the
number of random bits desired (Line~1). \mechanism~then prepares to generate
random numbers in Lines~2-6 by first selecting DRAM words (i.e., the
granularity at which \jk{a DRAM module} is accessed) containing known RNG cells
for generating random data (Line~3).  To maximize \jkfour{the} throughput of
random number generation, \mechanism~\mpy{chooses} DRAM words with the highest
density of RNG cells in each bank (to exploit DRAM parallelism). Since each
DRAM access can induce activation failures \emph{only} in the accessed DRAM
word, the density of RNG cells per DRAM word determines the number of random
bits D-RaNGe can generate per access.    
For each available DRAM bank, \mechanism~selects two DRAM words (in
distinct DRAM rows) containing RNG cells. The purpose of selecting two DRAM
words in \emph{different} rows is to \emph{repeatedly} cause \emph{bank conflicts}, or
\jk{issue} requests to \emph{closed} DRAM rows so that every read request will
\emph{immediately} follow an activation. This is done by alternating accesses
to the chosen DRAM words in different DRAM rows. After selecting DRAM words
for generating random values, \mechanism~ writes a known data pattern that
results in high entropy to each chosen DRAM word and its neighboring cells
(Line~4) and gains exclusive access to rows containing the two chosen DRAM
words as well as their neighboring cells (Line~5).\footnote{Ensuring
exclusive access to DRAM rows can be done by remapping rows to 1) redundant
DRAM rows or 2) buffers in the memory controller so that these rows are hidden
from the \jkfour{system software} and only accessible by the memory controller for
generating random numbers. }  This ensures that the data pattern surrounding
the RNG cell and the original value of the RNG cell stay constant prior to each
access such that the failure probability of each RNG cell remains reliable (as
observed to be necessary in Section~\ref{subsec:dpd}).  To begin generating
random data (i.e., sampling RNG cells), \mechanism~reduces the value of
$t_{RCD}$ (Line~6).  From every available bank \jk{(Line~7)},
\mechanism~generates random values in parallel (Lines~8-15).  Lines 8 and 12
indicate the commands to alternate accesses to two DRAM words in distinct rows
of a bank to both 1) induce activation failures and 2) precharge the
\hhtwo{recently-accessed} row.  After inducing activation failures in a DRAM
word, \mechanism~extracts the value of the RNG cells within the DRAM word
(Lines~9 and 13) to use as random data and restores the DRAM word to its
original data value (Lines~10 and 14) to maintain the original data pattern.
Line~15 ensures that writing the original data value is complete before
attempting to \jk{sample} the DRAM words again. Lines~16 and 17 simply end the
loop if enough random bits of data have been harvested.  Line~18 sets the
$t_{RCD}$ timing parameter back to its default value, so other applications can
access DRAM without corrupting data. Line~19 releases exclusive access to the
rows containing the chosen DRAM words and their neighboring rows. 

We find that this methodology maximizes the opportunity for activation failures
in DRAM, thereby maximizing the rate of generating random data from RNG
cells.

\subsection{Full System Integration} 
\label{subsec:full_system_integration} 

In this work, we focus on developing a flexible substrate for sampling RNG
cells fully from within the memory controller. D-RaNGe generates random
numbers using a simple firmware routine running entirely within the memory
controller.  The firmware executes the sampling algorithm
(Algorithm~\ref{alg:lat_rng}) whenever \mpx{an application requests random
samples and there is available DRAM} bandwidth (i.e., DRAM is not servicing other
\jk{requests} or maintenance commands). In order to minimize latency between
requests for samples and their \mpx{corresponding} responses, a small queue of
already-harvested random data may be maintained in the memory controller for
use by the system. Overall performance overhead can be minimized by tuning
\mpx{both 1) the queue size and 2) how the memory
controller prioritizes requests for random numbers relative to normal memory
requests}.

In order to integrate \mechanism~with the rest of the system, the system
designer needs to decide how to best expose an interface by which an
application can \jk{leverage \mechanism~to generate true random numbers} on
their system. There are many ways to achieve this, including, but not limited
to:
\begin{itemize}
\item \mpx{Providing a simple \texttt{REQUEST} and \texttt{RECEIVE} interface
for applications to request and receive the random numbers using memory-mapped
configuration status registers (CSRs)~\cite{wolrich2004mapping} or other existing I/O
datapaths (e.g., x86 \texttt{IN} and \texttt{OUT} opcodes, Local Advanced
Programmable Interrupt Controller (LAPIC configuration \cite{intel32intel}).}



\item \mpx{Adding a new ISA instruction (e.g., Intel
\texttt{RDRAND}~\cite{hamburg2012analysis}) that retrieves random numbers from
the memory controller and stores them into processor registers.}
\end{itemize} 
The operating system may then expose \mpx{one or more of these interfaces to
user applications through standard kernel-user interfaces (e.g., system calls,
file I/O, operating system APIs). The system designer has complete freedom to
choose between these (and other) mechanisms that expose an interface for user
applications to interact with D-RaNGe. We expect that the best option will be
\hhtwo{system specific}, depending both on the desired D-RaNGe use cases and the ease
with which the design can be implemented.}



\setstretch{0.82}
\section{\mechanism~Evaluation} 
\label{sec:evaluation} 

We evaluate \mpy{three key aspects of} \mechanism. \mpy{First, we} show that
the random data \jkthree{obtained from RNG cells identified by D-RaNGe}
\mpy{passes} all of the tests in the NIST test suite for randomness
(Section~\ref{subsec:nist}). Second, we analyze the
\jkthree{existence} of RNG cells across 59 LPDDR4 and 4 DDR3 DRAM chips (due to long
testing time) randomly sampled from the overall population of DRAM chips across
all three major DRAM manufacturers (Section~\ref{subsec:diff_chips}). Third, we
evaluate \mechanism~in terms of the six key properties of an ideal TRNG as
explained in Section~\ref{sec:motivation}
(Section~\ref{subsec:trng_key_char_eval}). 


\subsection{NIST Tests} 
\label{subsec:nist} 


First, we identify RNG cells using our RNG cell \jkthree{identification}
process (Section~\ref{subsec:cell_selection}). Second, we sample \jkfour{\emph{each}}
identified RNG \hhf{cell} \jkthree{one million times} to generate large amounts of
random data (\mpy{i.e.,} 1~Mb \emph{bitstreams}). Third, we evaluate the
entropy of the bitstreams from the identified RNG cells with the NIST test
suite for randomness~\cite{rukhin2001statistical}. Table~\ref{Tab:NIST} shows
the average results of 236 1~Mb bitstreams\footnote{We test data obtained from
4 RNG cells from each of 59 DRAM chips, to maintain a reasonable NIST testing
time and \jkfour{thus} show that RNG cells across all \jkfour{tested} DRAM
chips reliably generate random values.} across the 15 tests of the full NIST
\begin{table}[h!]
\footnotesize
\begin{center}
\begin{tabular}{ |c||c|c|c }
\cline{1-3}
\textbf{NIST Test Name} & \textbf{P-value} & \textbf{Status} \\
\cline{1-3}
\hhline{|=|=|=|}
monobit & 									0.675  & PASS \\ 
frequency\_within\_block & 					0.096  & PASS \\ 
runs & 										0.501  & PASS \\ 
longest\_run\_ones\_in\_a\_block & 			0.256  & PASS \\ 
binary\_matrix\_rank & 						0.914  & PASS \\ 
dft & 										0.424  & PASS \\ 
non\_overlapping\_template\_matching &	 	>0.999 & PASS \\ 
overlapping\_template\_matching & 			0.624  & PASS \\ 
maurers\_universal & 						0.999  & PASS \\ 
linear\_complexity & 						0.663  & PASS \\ 
serial & 									0.405  & PASS \\ 
approximate\_entropy & 						0.735  & PASS \\ 
cumulative\_sums & 							0.588  & PASS \\ 
random\_excursion & 						0.200  & PASS \\ 
random\_excursion\_variant & 				0.066  & PASS \\ 
\cline{1-3}
\end{tabular}
\caption{\hhf{\mechanism\ results} with NIST randomness test suite. \vspace{-10pt}}
\label{Tab:NIST} 
\end{center}
\end{table}
test suite for randomness. P-values are calculated for each test,\footnote{A
p-value close to 1 indicates that we must accept the null hypothesis, while a
p-value close to 0 and below a small threshold, e.g., $\alpha=0.0001$
(\jkthree{recommended by the NIST Statistical Test Suite
documentation~\cite{rukhin2001statistical}}), indicates that we must reject the
null hypothesis.} where the null hypothesis for each test is that a perfect
random number generator would \emph{not} have produced random data with
\emph{better} characteristics for the given test than the tested
sequence~\cite{marton2015interpretation}. Since the resulting P-values for each
test in the suite are greater than our chosen level of significance,
$\alpha=0.0001$, we accept our null hypothesis for each test. We note that all
236 bitstreams pass all 15 tests with similar P-values.  Given our
$\alpha=0.0001$, our proportion of passing sequences (1.0) falls within the
range of acceptable proportions of sequences that pass each test
(\jkthree{[0.998,1]} calculated by \mpy{the} NIST statistical test suite using
$(1-\alpha)\pm3\sqrt{\frac{\alpha(1-\alpha)}{k}}$, where $k$ is the number of
tested sequences). This \emph{strongly} indicates that \mechanism~can generate
unpredictable, \mpy{truly} random values. Using the proportion of 1s and 0s
generated from each RNG cell, we calculate \mpy{Shannon}
entropy~\cite{shannon1948mathematical} and find the \emph{minimum} entropy
across all RNG cells to be 0.9507.

\subsection{RNG Cell Distribution} 
\label{subsec:diff_chips} 

\begin{figure}[h] 
    \centering \includegraphics[width=\linewidth]{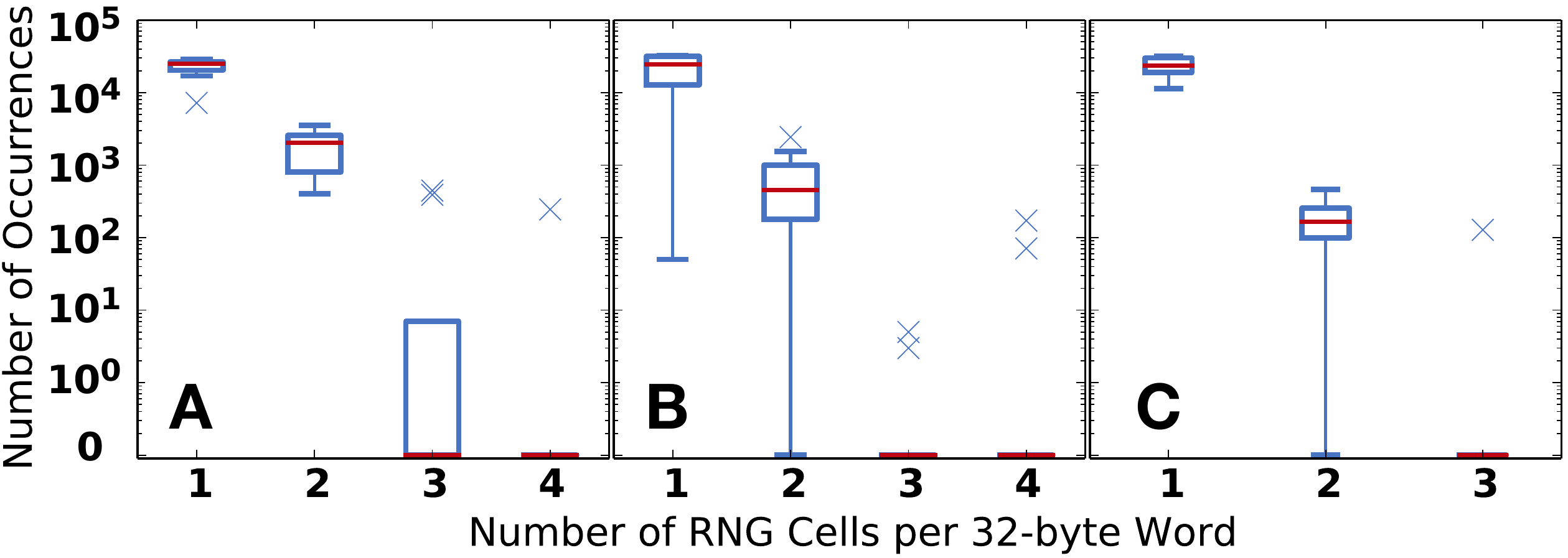} 
    \caption{Density of RNG cells in DRAM words per bank.} 
    \label{fig:bits_per_word} 
\end{figure}

The throughput at which \mechanism~generates random numbers is a function of
the 1) density of RNG cells per DRAM word and 2) bandwidth \jkthree{with which we
can access} DRAM words when using our methodology for inducing activation
failures. Since each DRAM access can induce activation failures \emph{only} in
the accessed DRAM word, the density of RNG cells per DRAM word indicates the
number of random bits D-RaNGe can sample per access.  We first study the
density of RNG cells per word across DRAM \jkthree{chips}.
Figure~\ref{fig:bits_per_word} plots the distribution of the number of words
containing x RNG cells (\jkthree{indicated by the value on the} x-axis) per
\emph{bank} across 472 banks from 59 DRAM devices from \jkthree{all manufacturers}.
The distribution is presented as a box-and-whiskers plot where the y-axis
\jkthree{has} a logarithmic scale with a 0 point. The three plots respectively
show the distributions for DRAM devices from the three manufacturers
(indicated at the bottom left corner of each plot). 

We make three key observations. First, RNG cells are \emph{widely available in
every bank} across many chips. This means that we can use the available DRAM
access parallelism that multiple banks offer and sample RNG cells from each
DRAM bank in parallel to improve random number generation throughput.  Second,
\emph{every} bank that we \mpy{analyze} has \emph{multiple DRAM words}
containing at least one RNG cell. The DRAM bank with the smallest
\mpy{occurrence} of RNG cells has 100 DRAM words containing only 1 RNG cell
(manufacturer B). Discounting this point, the distribution of the number of
DRAM words containing only 1 RNG cell is tight with a high number of RNG cells
(e.g., tens of thousands) in each bank, regardless of the manufacturer. Given
our random sample of DRAM chips, we expect that the existence of RNG cells in
DRAM banks will hold true for all DRAM chips. Third, we observe that a single
DRAM word can contain as many as 4 RNG cells. Because the throughput of
accesses to DRAM is fixed, the number of RNG cells in the accessed words
essentially acts as a multiplier for the throughput of random numbers
generated (e.g., accessing DRAM words containing 4 RNG cells results in
\emph{4x} the throughput of random numbers compared to accessing DRAM words
containing 1 RNG cell).

\subsection{TRNG Key Characteristics Evaluation} 
\label{subsec:trng_key_char_eval} 

We now evaluate \mechanism~in terms of the six key properties of an effective
TRNG as explained in Section~\ref{sec:motivation}. 

\textbf{Low Implementation Cost.} To induce activation failures, we must be
able to reduce the DRAM timing parameters below manufacturer-specified values.
Because memory controllers issue memory accesses according to the timing
parameters specified in a set of internal registers, D-RaNGe requires
\jkfour{simple} software support to be able to programmatically modify the
memory controller's registers.  Fortunately, there exist some
processors~\cite{lee2015adaptive, opteron_amd, bkdg_amd2013, ram_overclock}
that \emph{already} enable software to directly change memory controller
register values, i.e., the DRAM timing parameters. These processors can easily
generate random numbers with \mechanism. 

All other processors that do \emph{not} currently support direct changes to
memory controller registers require \emph{minimal} software changes to expose
an interface for changing the memory controller
registers~\cite{ARM2016memrefman, samsung2014refman, hassan2017softmc,
softmc-safarigithub}. To \jkthree{enable} a more efficient implementation, the
memory controller could be programmed such that it issues DRAM accesses with
distinct timing parameters on a per-access granularity to reduce the overhead
in 1) changing the DRAM timing parameters and 2) allow concurrent DRAM
accesses by other applications. In the rare case where these registers are
unmodifiable by even the hardware, the hardware changes necessary to enable
register modification \mpy{are} minimal and \mpy{are} simple to
implement~\cite{lee2015adaptive, hassan2017softmc, softmc-safarigithub}.

We experimentally find that we can induce activation failures \mpy{with}
$t_{RCD}$ between $6ns$ and $13ns$ (\mpy{reduced} from the default of $18ns$). Given
this wide range of failure-inducing $t_{RCD}$ values, most memory controllers
should be able to adjust their timing parameter registers to a value within
this range.

\textbf{Fully Non-deterministic.} As we have shown in
Section~\ref{subsec:nist}, the bitstreams extracted from the
\jkthree{D-RaNGe-identified} RNG cells pass \emph{all} 15 NIST tests\jkthree{. We have}
full reason to believe that we are inducing a metastable state of the sense
amplifiers (as hypothesized by~\cite{chang2016understanding}) such that we are
effectively sampling random physical phenomena to extract unpredictable random
values.  

\begin{figure}[h]
\centering \includegraphics[width=\linewidth]{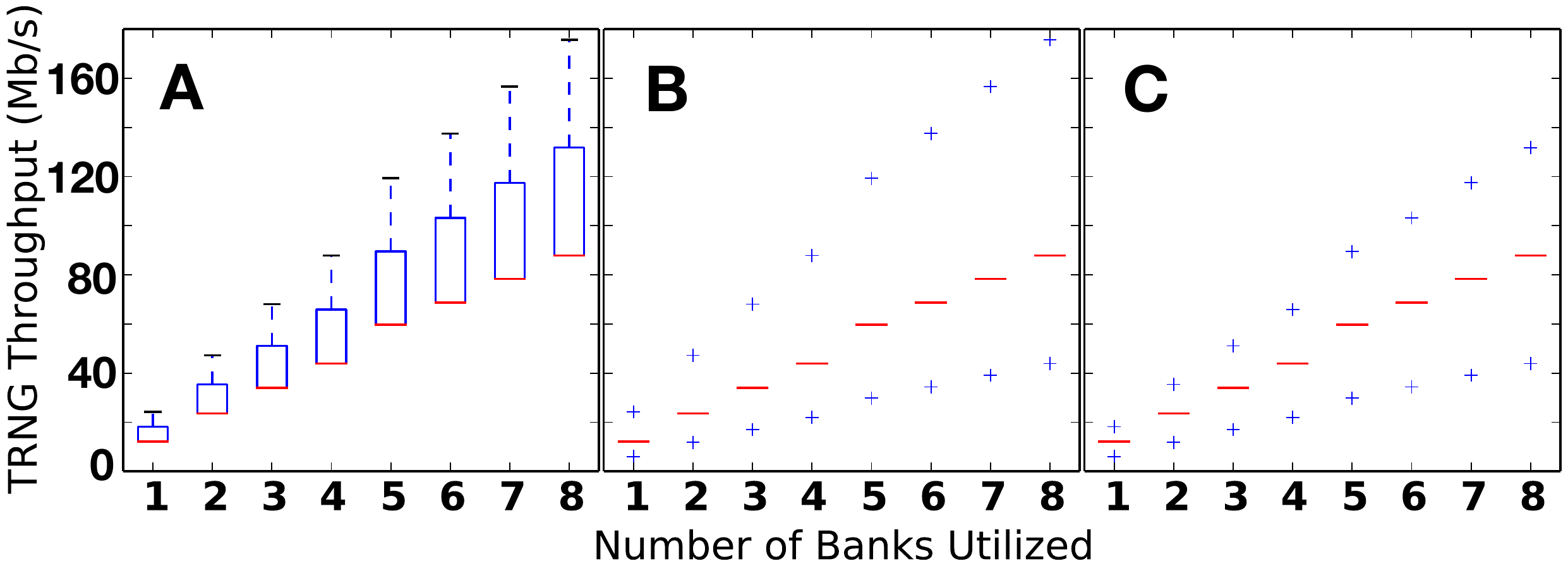} 
\caption{Distribution of TRNG throughput across chips.} 
\label{fig:trng_throughput}
\end{figure}

\textbf{High Throughput of Random Data.} Due to the various use cases of random
number generation discussed in Section~\ref{sec:motivation}, different
applications have different throughput requirements for random number
generation, and applications may tolerate a reduction in performance so
that \mechanism~can quickly generate true random numbers. Fortunately,
\mechanism~provides flexibility to tradeoff between the \textit{system
interference} it causes, i.e., the slowdown experienced by concurrently running
applications, and the random number generation throughput it provides. To
demonstrate this flexibility, Figure~\ref{fig:trng_throughput} plots the TRNG
throughput of \mechanism~when using varying numbers of banks (\emph{x} banks on
the x-axis) across the three DRAM manufacturers (indicated at the top left
corner of each plot). For each number of banks used, we plot the distribution
of TRNG throughput that we observe \emph{real} DRAM devices to provide\jkthree{. The
available density of RNG cells in a DRAM device (provided in
Figure~\ref{fig:bits_per_word}) dictates the TRNG throughput that the DRAM
device can provide. We plot each distribution} as
a box-and-whiskers plot. For each number of banks used, we select x
banks with the greatest sum of RNG cells across each banks' two DRAM words with
the highest density of RNG cells (that are \emph{not} in the same DRAM row).
We select two DRAM words per bank because we must alternate accesses between
two DRAM rows (as shown in \jkfour{Lines 8 and 12} of Algorithm~\ref{alg:lat_rng}).
The sum of the RNG cells available across the two selected DRAM words for each
bank is considered each bank's \emph{TRNG data rate}, and we use this value to
obtain \mechanism's throughput. We use Ramulator~\cite{ramulatorgithub,
kim2016ramulator} to obtain the rate at which we can execute the core loop of
Algorithm~\ref{alg:lat_rng} with varying numbers of banks. We obtain the random
number generation throughput for x banks with the following equation:   \vspace{-4pt}
\begin{equation} \label{eq:x_bank_throughput}
TRNG\_Throughput_{x\_Banks} = \sum_{n=1}^{x} \frac{TRNG\_data\_rate_{Bank\_n}}{Alg2\_Runtime_{x\_banks}} 
\end{equation} 
where $TRNG\_data\_rate_{Bank\_n}$ is the TRNG data rate for the selected bank,
and $Alg2\_Runtime_{x\_banks}$ is the runtime of the core loop of
Algorithm~\ref{alg:lat_rng} when using $x$ Banks. We note that because we
observe small variation in the density of RNG cells per word (between 0 and 4),
we see that \jkthree{TRNG throughput across different chips is} generally very similar.
For this reason, we see that the box and whiskers are condensed into a single
point for distributions of manufacturers B and C.  We find that when
\emph{fully} using \emph{all} 8 banks in a single DRAM channel, every
device can produce \emph{at least} 40 Mb/s of random data regardless of
manufacturer.  The highest throughput we observe from devices of manufacturers
A/B/C respectively are \changes{179.4/134.5/179.4 Mb/s}. On average, across all
manufacturers, we find that \mechanism~can provide a throughput of
\changes{108.9 Mb/s.} 

We draw two key conclusions. First, due to the \emph{parallelism} of multiple
banks, the throughput of random number generation increases linearly as we
use more banks. Second, there is variation of TRNG throughput across
different DRAM devices, but the medians across manufacturers are very 
similar. 

We note that any throughput sample point on this figure can be multiplied by
the number of available channels in a memory hierarchy for a better TRNG
throughput estimate for a system with multiple DRAM channels.  For an example 
memory hierarchy comprised of 4 DRAM channels, \mechanism~results in a maximum
(average) throughput of 717.4 Mb/s (435.7 Mb/s). 

\textbf{Low Latency.} \jkfour{Since D-RaNGe's sampling mechanism consists of a
single DRAM access, the latency of generating random values is directly related
to the DRAM access latency. Using the timing parameters specified in the JEDEC
LPDDR4 specification~\cite{2014lpddr4}, we calculate D-RaNGe's latency to
generate a 64-bit random value. To \hhf{calculate the} \emph{maximum latency} for D-RaNGe,
we assume that 1) each DRAM access provides only 1 bit of random data (i.e.,
each DRAM word contains \emph{only} 1 RNG cell) and 2) we can use only a single
bank within a single channel to generate random data. We find that D-RaNGe can
generate 64 bits of random data with a \emph{maximum} latency of $960 ns$. If
D-RaNGe takes full advantage of DRAM's \hhf{channel- and bank-level}
parallelism in a system with 4 DRAM channels \jkfive{and 8 banks per channel,} D-RaNGe
can generate 64 bits of random data by issuing 16 DRAM accesses per channel in
parallel. This results in a latency of $220 ns$.  To \hhf{calculate the}
\emph{empirical minimum latency} for D-RaNGe, we fully parallelize D-RaNGe
\hhf{across banks in all 4 channels} while \hhf{also} assuming that each DRAM
access provides 4 bits of random data, since we find a maximum density of 4 RNG
cells per DRAM word in the LPDDR4 DRAM devices that we characterize
(Figure~\ref{fig:bits_per_word}).  We find the empirical minimum latency to be
\emph{only} $100 ns$ in our tested devices.}


\textbf{Low System Interference.} The flexibility of using a different number
of banks across the available channels in a system's memory hierarchy allows
\mechanism~to \jkthree{cause} varying levels of system interference at the
expense of TRNG throughput. \jkfour{This enables application developers to
generate random values with D-RaNGe at varying tradeoff points} depending on
the running applications' \jkfour{memory access} requirements. We analyze
\mechanism's system interference with respect to DRAM storage overhead and DRAM
latency. 

In terms of storage overhead, \mechanism~simply requires exclusive access
rights to \jkfour{six DRAM rows} per bank, consisting of the two rows
containing the RNG cells and each row's two \jkfour{physically-adjacent} DRAM
rows containing the chosen data pattern.\footnote{As in prior
work~\cite{kim2014flipping, bains2015method, mutlu2017rowhammer}, we argue that
manufacturers can disclose \jkfour{which rows are physically adjacent to each
other.}} This results in an insignificant 0.018\% DRAM storage overhead cost. 

\jkthree{To evaluate} \mechanism's effect on \mpy{the} DRAM \jkthree{access
latency of regular memory requests}, we present one implementation of
\mechanism. For a single DRAM channel, which is the granularity at which DRAM
timing parameters are applied, \mechanism~can alternate between using a
reduced $t_{RCD}$ and the default $t_{RCD}$. When using a reduced $t_{RCD}$,
\mechanism~\mpy{generates} random numbers across every bank in the channel. On
the other hand, when using the default $t_{RCD}$, memory requests from running
applications \mpy{are} serviced to ensure application progress.  The length of
these time intervals (with default/reduced $t_{RCD}$) can both be adjusted
according to the applications' random number \jkthree{generation}
requirements.  Overall,
\mechanism~\jkthree{provides} significant flexibility in \jkthree{trading off} its system
overhead \jkthree{with its TRNG} throughput. \jkthree{However,} it is
up to the system designer to use and exploit the flexibility for their
requirements.  To show the potential throughput of \mechanism~without impacting
\jkthree{concurrently-running} applications, we run simulations with
\jkfour{the SPEC CPU2006}~\cite{spec2006} workloads, and calculate the idle DRAM bandwidth
available that we can use to issue D-RaNGe commands. We find \mpy{that,}
across all workloads, we can obtain an average (maximum, minimum) random-value
throughput of 83.1 (98.3, 49.1) Mb/s \mpy{with \emph{no} significant impact on
overall system performance}. 


\textbf{Low Energy Consumption.} To evaluate the energy consumption of D-RaNGe,
we use DRAMPower~\cite{drampowergithub} to analyze the output traces of
Ramulator~\cite{ramulatorgithub, kim2016ramulator} when DRAM is (1)
generating random numbers (Algorithm~\ref{alg:lat_rng}), and (2) idling and not
servicing memory requests. We subtract quantity (2) from (1) to
obtain the estimated energy consumption of D-RaNGe. We then divide the value by
the total number of random bits found during execution and find that, on
average, D-RaNGe finds random bits at the cost of 4.4 nJ/bit.


\setstretch{0.83} 
\section{Comparison with Prior DRAM TRNGs} 
\label{comparison}

To our knowledge, this paper provides the highest-throughput TRNG \hh{\emph{for
commodity DRAM devices}} by exploiting activation failures as a sampling
mechanism for observing entropy in DRAM cells. There are a number of proposals
to construct TRNGs using commodity DRAM devices, which \jkthree{we summarize} in
Table~\ref{tab:prior_dram_works} based on their entropy sources.  In this
section, we compare each of these works with \mechanism. \jkthree{We} show how
\mechanism\ fulfills the six key properties of an ideal TRNG
(Section~\ref{sec:motivation}) better than any prior DRAM-based TRNG proposal.
We group our comparisons by the entropy source of each prior DRAM-based TRNG
proposal.

\begin{table*}[h!]
\footnotesize
\begin{center}
\begin{tabular}{ |c||c|c|c|c|c|c|c| }
\hline 
      \textbf{Proposal}
	& \textbf{Year}
	& \begin{tabular}{@{}c@{}}\textbf{Entropy} \\ \textbf{Source}\end{tabular} 
	& \begin{tabular}{@{}c@{}}\textbf{True} \\ \textbf{Random}\end{tabular} 
	& \begin{tabular}{@{}c@{}}\textbf{Streaming} \\ \textbf{Capable}\end{tabular} 
	& \begin{tabular}{@{}c@{}}\textbf{\hhf{64-bit TRNG}} \\ \textbf{Latency}\end{tabular} 
	& \begin{tabular}{@{}c@{}}\textbf{Energy} \\ \textbf{Consumption}\end{tabular} 
	& \begin{tabular}{@{}c@{}}\textbf{Peak} \\ \textbf{Throughput}\end{tabular} \\ 
\hline \hline
Pyo+~\cite{pyo2009dram}
	& 2009
	& Command Schedule
	& \xmark  
	& \cmark  
    & $18{\mu}s$ 
    & N/A 
	& $3.40 Mb/s$ \\
\hline
Keller+~\cite{keller2014dynamic}          
	& 2014
	& Data Retention
	& \cmark  
	& \cmark  
    & $40 s$
    & $6.8mJ/bit$
	& $0.05 Mb/s$  \\
\hline
Tehranipoor+~\cite{tehranipoor2016robust} 
	& 2016
	& Startup Values
	& \cmark  
	& \xmark  
    & $>60ns$ \jkfour{(optimistic)} 
    & $>245.9 pJ/bit$ \jkfour{(optimistic)} 
	& N/A  \\
\hline
Sutar+~\cite{sutar2018d}                  
	& 2018
	& Data Retention
	& \cmark  
	& \cmark  
    & $40 s$ 
    & $6.8 mJ/bit$
	& $0.05 Mb/s$ \\
\hline
\textbf{\mechanism} 
	& 2018
	& Activation Failures
	& \cmark  
	& \cmark  
    & $100 ns < x < 960 ns$ 
    & $4.4 nJ/bit$
	& $717.4 Mb/s$ \\
\hline
\end{tabular}
\caption{Comparison to previous DRAM-based TRNG proposals. \vspace{-10pt}} 
\label{tab:prior_dram_works}
\end{center}
\end{table*}

\subsection{DRAM Command Scheduling} 

Prior work~\cite{pyo2009dram} proposes using non-determinism in DRAM command
scheduling for true random number generation. In particular, since pending
access commands contend with regular refresh operations, the latency of a DRAM
access is hard to predict and is useful for random number generation. 

Unfortunately, this method fails to satisfy two important properties of an
ideal TRNG. First, it harvests random numbers from the instruction and DRAM
command scheduling decisions made by the processor and memory controller, which
does \emph{not} constitute a fully non-deterministic entropy source. Since the
quality of the harvested random numbers depends directly on the quality of the
processor and memory controller implementations, the entropy source is visible
to and potentially modifiable by an adversary \hh{(e.g., by simultaneously
running a memory-intensive workload on another processor
core~\cite{moscibroda2007memory})}. Therefore, this method does not meet our
design goals as it does not securely generate random numbers. 

Second, although this technique has a higher throughput than those based on
DRAM data retention (Table~\ref{tab:prior_dram_works}), \mechanism\ still
outperforms this method in terms of throughput by 211x (maximum) and 128x
(average) because a single byte of random data requires a \emph{significant}
amount of time to generate. Even if we scale the throughput results provided
by~\cite{pyo2009dram} to a modern day system (e.g., $5 GHz$ processor, 4 DRAM
channels\footnote{The authors do not provide their DRAM configuration, so we
optimistically assume that they evaluate their proposal using one DRAM channel.
We also assume that by utilizing 4 DRAM channels, the authors can harvest four
times the entropy, which gives the benefit of the doubt
to~\cite{pyo2009dram}.}), the theoretical maximum throughput of \jkthree{Pyo et
al.'s} approach\footnote{We base our estimations on \cite{pyo2009dram}'s claim
that they can harvest one byte of random data every 45000 cycles. However,
using these numbers along with the authors' stated processor configuration
(i.e., $2.8 GHz$) leads to a discrepancy between our calculated maximum
throughput ($\approx 0.5 Mb/s$) and that reported in~\cite{pyo2009dram}
($\approx 5 Mb/s$). We believe our estimation methodology and calculations are
sound. In our work, we compare \mechanism's peak throughput against that of
\cite{pyo2009dram} using a more modern system configuration (i.e., $5 GHz$
processor, 4 DRAM channels) than used in the original work, which gives the
benefit of the doubt to~\cite{pyo2009dram}.} is \emph{only} $3.40 Mb/s$ as
compared with the maximum (average) throughput of $717.4 Mb/s$ ($435.7 Mb/s$)
for \mechanism. To calculate the latency of generating random values, we assume
the same system configuration with \hh{\cite{pyo2009dram}'s} claimed number of
cycles \hh{45000} to generate random bits. To provide 64 bits of random data,
\cite{pyo2009dram} takes 18$\mu$s, which is significantly higher than D-RaNGe's
\jkthree{minimum/maximum latency of $100ns/960ns$}. Energy
consumption for ~\cite{pyo2009dram} depends heavily on the entire system that
it is running on, so we do not compare against this \hh{metric}.


\subsection{DRAM Data Retention} 

Prior works~\cite{keller2014dynamic, sutar2018d} propose using DRAM data
retention failures to generate random numbers. Unfortunately, this approach is
\emph{inherently too slow} for high-throughput operation due to the long wait
times required to induce \jkthree{data retention failures in DRAM}. While the
failure rate can be increased by increasing the operating temperature, a wait
time on the order of seconds is required to induce \jkthree{enough}
failures~\cite{liu2013experimental, khan2014efficacy, qureshi2015avatar,
patel2017reaper, kim2018dram} to achieve \jkfour{high-throughput} random number
generation, which is orders of \jkfour{magnitude} slower than D-RaNGe. 

Sutar et al.~\cite{sutar2018d} report that they are able to generate {256-bit}
random numbers using a hashing algorithm (e.g., {SHA-256}) on a $4~MiB$ DRAM
block that contains \mpo{data retention errors resulting from having disabled
DRAM refresh} for 40~seconds. \mpf{Optimistically assuming a large DRAM
capacity of $32~GiB$ and ignoring the time required to read out and hash the
erroneous data, a waiting time of 40~seconds to induce data retention errors
allows for an estimated maximum random number throughput of $0.05~Mb/s$. This
throughput is already far \hh{smaller} than \mechanism's measured maximum
throughput of \changes{$717.4 Mb/s$}, and it would decrease linearly with
DRAM capacity.  Even if we were able to induce a large number of
data retention errors by waiting only 1 second, the \hh{maximum} random number
generation throughput would be $2~Mb/s$, i.e., orders of \jkfour{magnitude} 
\hh{smaller} than \jkthree{that of} \mechanism.}

Because \cite{sutar2018d} requires a wait time of 40~seconds before
producing any random values, its latency for random number generation is
extremely high \jkfour{(40s)}. \hh{In contrast,} D-RaNGe can produce random values
very quickly since it generates random values \jkthree{potentially with} each DRAM access
(10s of nanoseconds). D-RaNGe therefore has a latency many orders of \jkfour{magnitude} 
lower than \hh{Sutar et al.'s mechanism~\cite{sutar2018d}.} 

We estimate the energy consumption of retention-time \jkthree{based TRNG} mechanisms
with Ramulator\hh{~\cite{kim2016ramulator, ramulatorgithub}} and
DRAMPower\hh{~\cite{drampowergithub, chandrasekar2011improved}}. We model first
\mpy{writing data to} a 4MiB DRAM \jkfour{region} (to constrain the energy consumption
estimate to the region of interest), waiting for 40 seconds, and then reading
from that region. We then divide the energy consumption of these operations by
the number of bits found (256 bits). \hh{We find that the energy consumption
is} around $6.8 mJ$ per bit\hh{, which} is orders of \jkfour{magnitude} more costly
than \jkthree{that of} \mechanism, which provides random \jkthree{numbers} at
$4.4 nJ$ per bit.

\subsection{DRAM Startup Values} 

Prior work~\cite{tehranipoor2016robust, eckert2017drng} proposes using DRAM
startup values as random numbers. Unfortunately, this method is unsuitable for
continuous high-throughput operation since it requires a DRAM power cycle in
order to obtain random data. We are unable to accurately model the latency of
this mechanism since it relies on the startup time of DRAM (i.e., bus frequency
\hh{calibration}, temperature \hh{calibration}, timing register
initialization~\cite{ddr4operationhynix}). This \jkthree{heavily depends} on
the implementation of the system and \jkfour{the} DRAM device in use. Ignoring
these components, \hh{we estimate the throughput of generating random numbers
using startup values by taking into account only the latency of a single} DRAM
read (\emph{after} all initialization is complete), which \jkthree{is $60ns$}.
We model energy consumption ignoring the initialization phase as well, by
modeling the energy to read a MiB of DRAM and \hh{dividing} that quantity by
\cite{tehranipoor2016robust}'s claimed number of random bits found in that
region (420Kbit). \hh{Based on this calculation, we estimate energy consumption
as} $245.9pJ$ per bit. While \hh{the energy consumption of
\cite{tehranipoor2016robust}} is smaller than the energy cost of \mechanism, we
note that \hh{our energy estimation for \cite{tehranipoor2016robust}} does
\emph{not} account for the energy consumption required for initializing DRAM to
be able to read out the random values.  Additionally,
~\cite{tehranipoor2016robust} requires a full system reboot which is often
impractical for applications and for effectively providing a \emph{steady
stream of random values}. \cite{eckert2017drng} suffers from the same issues
since it uses the same mechanism as \cite{tehranipoor2016robust} to
generate random numbers and is strictly worse since \cite{eckert2017drng}
results in $31.8x$ less entropy. 

\subsection{Combining DRAM-based TRNGs} 

We note that \mechanism's method for sampling random values from DRAM is
entirely distinct from prior DRAM-based TRNGs that we have discussed in this
section. This makes it possible to combine \mechanism~with prior work to
produce random values at an even higher throughput.

\setstretch{0.81}
\section{Other Related Works} 
\label{related}


In this work, we focus on the design of a DRAM-based hardware mechanism to
implement a TRNG, which makes the focus of our work orthogonal to those that
design PRNGs. In contrast to prior DRAM-based TRNGs discussed in
Section~\ref{comparison}, we propose using \emph{activation failures} as an
entropy source. Prior works characterize activation failures in order to
exploit the resulting error patterns for overall DRAM latency
reduction~\hh{\cite{chang2016understanding, kim2018solar, lee-sigmetrics2017,
lee2015adaptive}} and to implement physical unclonable functions
(PUFs)~\cite{kim2018dram}.  However, none of these works measure the randomness
inherent in activation failures or propose using them to generate random
numbers.  

\hh{Many} TRNG designs have been proposed that exploit sources of entropy
\mpf{that are \emph{not}} based on DRAM. \mpf{Unfortunately, these proposals
either 1) require custom hardware modifications that preclude their application
to commodity devices, or 2) do not sustain continuous (i.e., constant-rate)
high-throughput operation}. We briefly discuss different entropy sources with
examples. 

\textbf{Flash Memory Read Noise.} Prior proposals use random
telegraph noise in flash memory devices as an entropy source (up to 1
Mbit/s)~\cite{wang2012flash, ray2018true}. Unfortunately, flash memory is
orders of magnitude slower than DRAM, making flash unsuitable for
high-throughput \jkthree{and low-latency} operation.

\textbf{SRAM-based Designs.} SRAM-based TRNG designs exploit
randomness in startup values \cite{ 
	  holcomb2007initial, holcomb2009power, 
	van2012efficient}. 
Unfortunately, these proposals are unsuitable for continuous, high-throughput
operation since they require a power cycle.

\textbf{GPU- and FPGA-Based Designs.}  Several works harvest random numbers
from GPU-based (up to 447.83 Mbit/s)~\cite{chan2011true, tzeng2008parallel,
teh2015gpus} and FPGA-based (up to 12.5 Mbit/s)~\cite{majzoobi2011fpga,
wieczorek2014fpga, chu1999design, hata2012fpga} entropy sources. \hh{These}
\mpf{proposals do not require modifications to commodity GPUs or FPGAs\hh{.
Yet,} GPUs and FPGAs are not as prevalent as DRAM in commodity devices
today.}

\textbf{Custom Hardware.} Various works propose TRNGs based in part or fully on
non-determinism \jkthree{provided by custom hardware designs} (\jkfour{with
TRNG throughput} up to 2.4 Gbit/s)~\cite{ 
	amaki2015oscillator, 
	yang2016all, 
	bucci2003high, 
	bhargava2015robust, 
	petrie2000noise, 
	mathew20122, brederlow2006low, tokunaga2008true, 
	kinniment2002design, holleman20083, holcomb2009power, 
	pareschi2006fast, 
	stefanov2000optical}.  
Unfortunately, \jkthree{the need for custom hardware limits the widespread use of such
proposals in} commodity hardware devices (today).
\delete{\textbf{System-Level RNGs.} Operating systems often provide interfaces
for harvesting entropy from devices running on a system.  Unfortunately, the
entropy source of a system-level RNG depends on 1) the particular devices
attached to the system (e.g., I/O peripherals, disk drives) and 2) the quality
of the drivers that directly harvest entropy from these devices. This means
that a system-level RNG is not guaranteed to be harvesting random numbers from
a fully non-deterministic entropy source; therefore, system-level RNGs do not
meet our design goals.}

\vspace{-8pt}
\setstretch{0.81}
\section{Conclusion}
\label{sec:conclusion} 

We propose \mechanism, a mechanism for extracting true random numbers with high
throughput from unmodified commodity DRAM devices on any system that allows
manipulation of DRAM timing parameters in the memory controller.  \mechanism\
harvests fully non-deterministic random numbers from DRAM row activation
failures, which are \jk{bit errors} induced by intentionally \jk{accessing DRAM
with lower latency than required for correct row activation.} Our TRNG is based
on two key observations: 1) activation failures can be induced quickly and 2)
repeatedly accessing certain DRAM cells with reduced activation latency results
in reading true random data. We validate the quality of our TRNG with the
commonly-used NIST statistical test suite for randomness. Our evaluations show
that \mechanism\ significantly outperforms the previous highest-throughput
DRAM-based TRNG by up to 211x \jkthree{(128x on average)}. We conclude that
DRAM row activation failures can be effectively exploited to efficiently
generate true random numbers with high throughput on a wide range of devices
that use commodity DRAM chips.

\setstretch{0.82}


\section*{Acknowledgments} \hh{We} thank the anonymous reviewers \jk{of
HPCA 2019 and MICRO 2018} for feedback and the SAFARI group members for
feedback and the stimulating intellectual environment they provide.



%


\SetTracking
 [ no ligatures = {f},
 outer kerning = {*,*} ]
 { encoding = * }
 { -40 } 

{

  \let\OLDthebibliography\thebibliography
  \renewcommand\thebibliography[1]{
    \OLDthebibliography{#1}
    \setlength{\parskip}{0pt}
    \setlength{\itemsep}{0pt}
  }
  \bibliographystyle{IEEEtranS}
  \bibliography{ref}
}

\end{document}